\documentclass[smallextended]{svjour3}       
\smartqed  
\usepackage{graphicx}
\usepackage{epstopdf}
\usepackage{amsmath}
\usepackage{amssymb}
\usepackage{hhline}
\usepackage{enumfrom}
\usepackage{enumerate}
\usepackage{color}
\usepackage{multicol}
\newcommand{\red}[1]{#1}

\newcommand{\ignore}[1]{}

\newcommand{\boxtheorem}{\hfill $\Box$}
\newcommand{\nit}[1]{{\it #1}}
\newcommand{\mc}[1]{\mathcal{ #1}}

\newcommand{\mf}[1]{\mathfrak{ #1}}
\newcommand{\bcq}{BCQ}
\newcommand{\cq}{CQ}

\newcommand{\fo}{FO}
\newcommand{\n}{~{\it not}~}
\newcommand{\sfd}{{\sf d}}
\newcommand{\sfs}{{\sf s}}
\newcommand{\reps}{\nit{Rep}^{\!{\sf S}\!}(D,\kappa(Q))}
\newcommand{\nn}{\nit{null}}
\newcommand{\sfS}{{\sf S}}

\newcommand{\au}{\mathbf{u}}
\newcommand{\bu}{\mathbf{fu}}
\newcommand{\bft}{\mathbf{t}}
\newcommand{\s}{\mathbf{s}}
\newcommand{\St}{\nit{Store}}
\newcommand{\R}{\nit{Receives}}
\newcommand{\Rd}{\nit{Resold}}
\newcommand{\Py}{\nit{Pricey}}
\newcommand{\Q}{\nit{Supplied}}

\newcommand{\blue}[1]{\textcolor{blue}{#1}}
\newcommand{\comlb}[1]{{\vspace{2mm}\noindent \bf  {\blue{COMM(LEO):}}}~ #1 \hfill {\bf
    END.}\\}

\begin{document}

\title{Specifying and Computing Causes for Query Answers  in Databases via Database Repairs and Repair-Programs
}

\titlerunning{Specifying  Causes for Query Answers}        

\author{Leopoldo Bertossi}

\authorrunning{L. Bertossi} 

\institute{   {\bf Universidad Adolfo Ib\'a\~nez, Faculty of Engineering and Sciences, Santiago, Chile; and ``Millenium Institute on Foundations of Data" (IMFD, Chile)}. \\
              \email{leopoldo.bertossi@uai.cl}           
}

\date{Received: date / Accepted: date}

\maketitle

\begin{abstract}There is a recently established correspondence between database tuples as causes for query answers in databases and tuple-based repairs of inconsistent databases with respect to denial constraints.  In this work, answer-set programs that specify database repairs are used as a basis for solving computational and reasoning problems around  causality in databases, including causal responsibility. Furthermore, causes are introduced also at the attribute level by appealing to an attribute-based repair semantics that uses null values. Corresponding repair-programs are introduced, and used as a basis for computation and reasoning about attribute-level causes. The answer-set programs are extended in order to capture causality under integrity constraints. 

\keywords{Causality \and databases \and repairs \and  constraints \and answer-set programming}
\end{abstract}

\ignore{++
\section{Introduction}
\label{intro}
Your text comes here. Separate text sections with
\section{Section title}
\label{sec:1}
\subsection{Subsection title}
\label{sec:2}
as required. Don't forget to give each section
and subsection a unique label (see Sect.~\ref{sec:1}).
\paragraph{Paragraph headings} Use paragraph headings as needed.
\begin{equation}
a^2+b^2=c^2
\end{equation}

\begin{figure}
  \includegraphics{example.eps}
\caption{Please write your figure caption here}
\label{fig:1}       
\end{figure}
%
\begin{figure*}
  \includegraphics[width=0.75\textwidth]{example.eps}
\caption{Please write your figure caption here}
\label{fig:2}       
\end{figure*}
%
\begin{table}
\caption{Please write your table caption here}
\label{tab:1}       
\begin{tabular}{lll}
\hline\noalign{\smallskip}
first & second & third  \\
\noalign{\smallskip}\hline\noalign{\smallskip}
number & number & number \\
number & number & number \\
\noalign{\smallskip}\hline
\end{tabular}
\end{table}
++}

\section{ \ Introduction}

 Causality appears at the foundations of many scientific disciplines. In data and knowledge management,  causality may be related to some
 form of
 uncertainty about obtained results. For example, about why certain  query answers are obtained or not; or why certain semantic conditions are not satisfied. These tasks become more prominent and difficult when dealing with large volumes of data.
One would expect the database to provide {\em explanations}, which could be used to understand, explore and make sense of the data, or to reconsider
queries and integrity constraints (ICs). Causes for data phenomena can be seen as  explanations.

Building on work on {\em actual causality}, mainly developed in the context of artificial  intelligence \cite{Halpern05,Chockler04}, {\em causality in databases} was introduced in \cite{Meliou2010a}, where the notions of counterfactual intervention and structural model are applied. More specifically, \cite{Meliou2010a} introduces the notions of: (a) a database tuple as an {\em actual cause} for a query result, \ (b) a
{\em contingency set} for a cause, as a set of tuples that must accompany the cause for it to be such, and (c) the {\em responsibility} of a cause as a numerical measure of its strength.

\red{\begin{example}  \label{ex:cause0}  Consider the  relational database below, with table $\St$ representing official stores, and table $\R$, for stores receiving goods from other stores:
\begin{center}
\begin{tabular}{l|c|c|} \hline
$\R$  & $R.1$ & $R.2$ \\\hline
& $s_2$ & $s_1$\\
& $s_3$ & $s_3$\\
 & $s_4$ & $s_3$\\
 \hhline{~--}
\end{tabular} \hspace*{0.5cm}\begin{tabular}{l|c|c|}\hline
$\St$  & $S.1$  \\\hline
& $s_2$ \\
& $s_3$ \\
 & $s_4$ \\
\hhline{~-}
\end{tabular}
\end{center}
\noindent The tables include attribute names.  For accounting purposes, a store could be its own supplier, as shown by the tuple $\R(s_3,s_3)$. \ The
query asking if there are pairs of official stores in a receiving relationship is true in the database. The question is about the tuples that cause this query to be true, and how strong they are as causes. In this case, we would expect the   tuples $\R(s_3, s_3)$ and $\R(s_4, s_3)$ to be causes.   \boxtheorem
\end{example} }

Most of our research on causality in databases has been motivated by an attempt to understand causality in data and knowledge management from different perspectives, and profiting from the established connections between them. In \cite{tocs},
precise reductions between causality in databases, database repairs, and consistency-based diagnosis were established; and the relationships were investigated and exploited. In \cite{flairsExt}, causality in databases was related to view-based database updates and abductive diagnosis.
These are all interesting and fruitful connections among several forms of non-monotonic reasoning; each of them
reflecting  some form of  uncertainty about the information at hand. In the case of database  repairs \cite{Bertossi2011}, uncertainty due the non-satisfaction of given ICs, and it is represented by a class of possible repairs of the  inconsistent database.

Database repairs can be specified by means of {\em answer-set programs} (or {\em disjunctive logic programs with stable model semantics}) \cite{asp,gelfond,torsten}, the so-called {\em repair-programs}. Cf. \cite{monica,Bertossi2011} for details on repair-programs and additional references. In this work we exploit the reduction from database causality to database repairs established in \cite{tocs}, by taking advantage of repair-programs for
specifying and computing causes, their contingency sets, and their responsibility degrees.
We show that  the resulting {\em causality-programs} have the necessary and sufficient expressive power to capture and compute not only causes, which can be done with less expressive programs \cite{Meliou2010a}, but also minimal contingency sets and responsibilities, which provably requires higher expressive power. Causality programs can also be used for reasoning about causes.

As a finer-granularity alternative to tuple-based causes, we introduce a particular form of {\em attribute-based causes}, namely {\em null-based causes}, capturing the intuition that an attribute value may be the cause for a query to become true in the database. This is done by profiting from an abstract reformulation of the above mentioned relationship between tuple-based causes and tuple-based repairs. More specifically, we appeal to {\em null-based repairs}, which are a particular kind of {\em attribute-based repairs}.  According to null-based repairs, the inconsistencies of a database are solved by minimally replacing attribute values in tuples by (a properly formalized version of) {\footnotesize {\sf NULL}}, the null-value used in  SQL databases (with an SQL semantics). We also define the corresponding attribute-based notions of contingency set and responsibility.
We introduce repair (answer-set) programs for null-based repairs, so that the newly defined causes can be computed and reasoned about.

Finally, we briefly show how causality-programs can be adapted to give an account of other forms of causality in databases that are connected to other possible repair-semantics for databases.

More specifically, we make the following contributions:
\begin{enumerate}
\item We start from a characterization of
 actual causes for query answers in terms of minimal repairs that are based on tuple deletions from a database that does not satisfy certain denial constraints. Next, we propose an abstract notion of actual cause that depends on an abstract repair semantics.
 \item The abstract notion of cause is specialized by appealing to minimal repairs that are obtained through changes of attribute values by  {\footnotesize {\sf NULL}}, a null value that is treated as in SQL databases. In this way, we introduce a notion of actual cause at the attribute level (as opposed to tuple level, as is usually the case).\footnote{We appeal to an SQL semantics for null values, but this not crucial. What matters is that a null value cannot be used to satisfy a join or make true a comparison that uses a built-in predicate. Any other semantics that is compatible with these assumptions could be used instead.}

 \item We present answer-set programs (ASPs) for the specification and computation of causes and their responsibilities. They are extensions of repair ASPs, both at the tuple- and the attribute-level. In particular, we show how extensions of ASPs with sets, aggregations and weak program constraints can be used  for the computation of maximum-responsibility actual causes.

  \item   \red{The ASPs can be easily modified to accommodate {\em endogenous} and {\em exogenous} tuples, as considered under actual causality \cite{Meliou2010a}. The former are  somehow under our control, and can be  subjected to further analysis. Only they can be actual causes. The latter are external tuples, beyond our control, and taken as given. They are not considered as possible causes or contingent companions of causes. The notion of database repair with these two classes of tuples was introduced in \cite{tocs}. Writing and using ASPs for this kind of repairs is a rather straightforward extension of the programs we consider in this work, where all tuples are considered to be endogenous.}

 \item We show several  ASPs as examples, and their execution with the DLV and DLV-Complex systems \cite{dlv,calimeri08,calimeri09}.
  \item We elaborate on the notion of actual cause under given integrity constraints, and show how they can be computed via ASPs.
  \item We introduce several topics for relevant discussions and further research.
  \end{enumerate}

\red{The results obtained in this work are significant for several reasons. First of all, declarative specifications of causes in databases enable {\em logical reasoning} about causes and their responsibilities in combination with data. This can be done at the same level of the data, which, as facts,  become elements of an answer-set program. Using an ASP system such as  DLV  makes this possible without having to export the data as facts outside the database. After having an ASP taking care of the base specification of causes, one can consider combining  it with additional domain knowledge or semantic information, in the form of additional rules or constraints, or in combination with ontologies \cite{dlProgs}. One can also rather easily perform some forms of hypothetical reasoning about causes, by almost dynamically activating or disabling database tuples, or making some of them, or classes of them, {\em exogenous}, which means they cannot be considered as causes, and, on the repair side, cannot be considered for restoring consistency of the database.}

\red{Furthermore, the ASP paradigm is expressive enough to solve the intrinsically complex algorithmic tasks behind responsibility computation, in particular, of most responsible causes, without overkilling the problem. ASPs have exactly the required computational complexity for these tasks (c.f. Section \ref{sec:disc}). In this same direction, ASP systems have become particulary efficient, and are being used for solving hard combinatorial problems. In our case, we could do the actual computations directly using the ASP engines, without having to export the data to some external environment where {\em ad hoc} algorithms are implemented and run. Finally, the ASPs we propose can be used to specify an {\em abstract notion of cause} (c.f. Section \ref{sec:abstract}), from which specific forms of causes, such as tuple- and attribute-based causes can be obtained; and the generic ASPs can be easily adapted for those specific forms of causes, and others.}

This paper is structured as follows. Section \ref{sec:back} provides background material on relational databases, database causality, database repairs, and answer-set programming. Section \ref{sec:cauRepair} establishes correspondences between causes and repairs, and introduces an abstract notion of cause on the basis of an abstract repair-semantics.  Section \ref{sec:arepairs} introduces  null-based repairs and causes.  Section  \ref{sec:specTuples} presents repair-programs  for tuple-based causality computation and reasoning. Section  \ref{sec:abc} presents answer-set programs for null-based repairs and null-based causes. Section \ref{sec:ics} introduces actual causes in the presence of ICs, and illustrates the corresponding repair-programs that can be used for causality computation. Finally, Section \ref{sec:disc}, in more speculative terms,  contains a discussion about  research subjects that would naturally extend this work. \red{Appendices \ref{sec:dlv} and \ref{sec:B} show ASPs written and run in DLV.}  This paper is a revised and extended version of \cite{foiks18}.\footnote{This version contains a reformulated definition and presentation of null-based repairs, of attribute-based causes, several examples with programs in DVL, and the treatment of causes under integrity constraints.}  Its extended version \cite{foiks18Corr} contains additional examples with DLV.

\section{ \ Background}\label{sec:back}

\subsection{Relational databases} \vspace{-2mm}
  A relational schema $\mc{R}$ contains a domain, $\mc{C}$, of constants and a set, $\mc{P}$, of  predicates of finite arities. $\mc{R}$ gives rise to a language $\mf{L}(\mc{R})$ of first-order (FO)  predicate logic with built-in equality, $=$.  Variables are usually denoted by $x, y, z, ...$, and sequences thereof by $\bar{x}, ...$; and constants with $a, b, c, ...$, and sequences thereof by $\bar{a}, \bar{c}, \ldots$. An {\em atom} is of the form $P(t_1, \ldots, t_n)$, with $n$-ary $P \in \mc{P}$   and $t_1, \ldots, t_n$ {\em terms}, i.e. constants,  or variables.
  An atom is {\em ground}, a.k.a. a {\em tuple}, if it contains no variables. Tuples are denoted with $\tau, \tau_1, \ldots$. A database  instance, $D$, for schema $\mc{R}$ is a finite set of ground atoms; and it serves as a (Herbrand)  interpretation \ignore{The {\em active domain} of a database instance $D$, denoted ${\it Adom}(D)$, is the set of constants that appear in atoms of $D$.} structure for  language  $\mf{L}(\mc{R})$ \cite{lloyd} (cf. also Section \ref{sec:dasps}).

A {\em conjunctive query} (\cq)  is a \fo \ formula of the form \ $\mc{Q}(\bar{x})\!: \ \exists  \bar{y}\;(P_1(\bar{x}_1)\wedge \dots \wedge P_m(\bar{x}_m))$,
 with $P_i \in \mc{P}$, and (distinct) free variables $\bar{x} := (\bigcup \bar{x}_i) \smallsetminus \bar{y}$. If $\mc{Q}$ has $n$ free variables,  $\bar{c} \in \mc{C}^n$ \ is an {\em answer} to $\mc{Q}$ from $D$ if $D \models \mc{Q}[\bar{c}]$, i.e.  $Q[\bar{c}]$ is true in $D$  when the variables in $\bar{x}$ are componentwise replaced by the values in $\bar{c}$. $\mc{Q}(D)$ denotes the set of answers to $\mc{Q}$ from $D$. $\mc{Q}$ is a {\em Boolean conjunctive query} (\bcq) when $\bar{x}$ is empty. When it is {\em true} in $D$, by definition $\mc{Q}(D) := \{\nit{true}\}$. Otherwise, if it is {\em false}, $\mc{Q}(D) := \emptyset$. \ A {\em view} is predicate defined by means of a query, whose contents can be computed, if desired, by computing all the answers to the defining query.

 \begin{example}  \label{ex:cause00} \ (ex. \ref{ex:cause0} cont.) \ The  relational schema contains two predicates,  $\mc{R}=\{\St(\cdot),$  $ \R(\cdot,\cdot)\}$.
 A database $D$ compatible with this schema is shown in Example \ref{ex:cause0}. \ We will usually present a relational database as the set of its true atoms, i.e. the tuples in the tables with their predicates. In this case,
\ $D = \{\R(s_4,s_3),$ $ \R(s_2,s_1),$  $ \R(s_3,s_3), \St(s_4),$ $ \St(s_2),$ $ \St(s_3)\}.$

The query in Example \ref{ex:cause0}, asking as to whether there are official stores, such one receives goods from another, is a BCQ that can be written as

\centerline{$\mc{Q}\!: \ \exists x \exists y ( \St(x) \land \R(x, y) \land \St(y)).$}

\noindent  The query  is true in $D$, denoted \ $D \models \mc{Q}$.
\boxtheorem
\end{example}

In this work we consider integrity constraints (ICs), i.e. sentences of $\mf{L}(\mc{R})$,  that are: (a) {\em denial constraints} \ (DCs), i.e.  of the form

\centerline{$\kappa\!:  \neg \exists \bar{x}(P_1(\bar{x}_1)\wedge \dots \wedge P_m(\bar{x}_m))$}

 \noindent
where $P_i \in \mc{P}$, and $\bar{x} = \bigcup \bar{x}_i$; and (b) {\em functional dependencies} \ (FDs), i.e. of the form

\centerline{$\varphi\!:  \neg \exists \bar{x} (P(\bar{v},\bar{y}_1,z_1) \wedge P(\bar{v},\bar{y}_2,z_2) \wedge z_1 \neq z_2)$.}

\noindent Here,
$\bar{x} = \bar{y}_1 \cup \bar{y}_2 \cup \bar{v} \cup \{z_1, z_2\}$, and $z_1 \neq z_2$ is an abbreviation for $\neg z_1 = z_2$.\footnote{The variables in the atoms do not have to occur in the indicated order, but their positions should be in correspondence in the two atoms.}

An {\em inclusion dependency} is an IC of the form

\centerline{ $\psi\!: \ \forall \bar{x}(P(\bar{x}) \rightarrow \exists \bar{y}S(\bar{x}',\bar{y}))$,}

\noindent with $P, S \in \mc{P}$, and $\bar{x}' \subseteq \bar{x}$.

\begin{example}  \ \red{(ex. \ref{ex:cause00} cont.) \ For schema $\mc{R}$, the following is a denial constraint:}

\centerline{$\kappa:  \ \neg \exists x\exists y( \St(x)\wedge \R(x, y)\wedge \St(y)),$}

\noindent and the following is a functional dependency, of the second attribute upon the first, on predicate $\R$:

\centerline{$\varphi: \ \neg \exists x \exists y \exists z(\R(x,y) \wedge \R(x,z) \wedge y\neq z)$.}

The following is an inclusion dependency:

\centerline{$\psi\!: \ \forall x \forall z(\St(x) \rightarrow \exists y \R(x,y))$}

The constraint $\kappa$ is not satisfied by $D$, but $\varphi$ and $\psi$ are. \boxtheorem
\end{example}

\ignore{
A {\em key constraint} \ (KC) is a conjunction of FDs: \  $\bigwedge_{j=1}^k \neg \exists \bar{x} (P(\bar{v},\bar{y}_1) \wedge P(\bar{v},\bar{y}_2) \wedge y_1^j \neq y_2^j)$,
with $k = |\bar{y_1}| = |\bar{y}_2|$. }
\ A given schema may come with its set of ICs, and its instances are expected to satisfy them. If an instance does not satisfy them, we say it is {\em inconsistent}. In this work we concentrate mostly on DCs. \ See  \cite{ahv} for more details and background material on relational databases.

\subsection{Causality in databases} \label{sec:tcause}

A notion of {\em cause} as an explanation for a query result was introduced  in \ \ \cite{Meliou2010a}, as follows. For a
relational instance  ${D=D^n \cup D^x}$,  where ${D^n}$ and ${D^x}$ denote the mutually exclusive sets of {\em endogenous}  and {\em exogenous} tuples,
a tuple  ${\tau \in D^n}$ is called a
{\em counterfactual cause} for  a BCQ ${\mc{Q}}$,  if \ ${D\models \mc{Q}}$ \ and \ ${D\smallsetminus \{\tau\}  \not \models \mc{Q}}$. Now,
${\tau \in D^n}$ is an {\em actual cause} for  ${\mc{Q}}$
if there  exists ${\Gamma \subseteq D^n}$, called a {\em contingency set} for $\tau$,  such that ${\tau}$ is a  counterfactual cause for ${\mc{Q}}$ in ${D\smallsetminus \Gamma}$.
This definition is based on \cite{Halpern05}. 

The notion of {\em responsibility} reflects the relative degree of causality of a tuple for
a query result \cite{Meliou2010a} (based on \ \cite{Chockler04}).
 The { responsibility} of an actual cause ${\tau}$ for ${\mc{Q}}$, is ${\rho(\tau) \ := \ \frac{1}{|\Gamma| + 1}}$, where
${|\Gamma|}$ is the
size of a smallest contingency set for ${\tau}$. If $\tau$ is not an actual cause, $\rho(\tau):= 0$.
 Intuitively, tuples with { higher responsibility} provide stronger explanations.

The partition of the database into endogenous and exogenous tuples. Exogenous tuples are accepted as given, which may happen because we trust them, or we have little control on them, or are obtained from an external, trustable and indisputable data source, etc. Endogenous tuples are subject to experimentation and questioning, in particular, about their role in query answering or violation of ICs. The partition is application dependent, and we may not even have exogenous tuples. \ {\em Actually, in the following we will assume all the tuples in a database instance are endogenous}. \ (Cf. \cite{tocs} for the general case, and Section \ref{sec:disc} for additional discussions.)

 The notion of cause as defined above can be equally applied to answers to open CQs, say a cause for obtaining $\bar{a}$ as an answer to a CQ $\mc{Q}(\bar{x})$, i.e. a cause for $D \models \mc{Q}[\bar{a}]$.
Actually, it can be applied to monotonic queries in general, i.e. whose
sets of answers may only grow when the database grows \cite{tocs}. For example, CQs, unions of CQs (UCQs) and Datalog queries are monotonic. Causality for these queries was investigated in \cite{tocs,flairsExt}. \ In this work we concentrate mostly on conjunctive queries, possibly with built-in comparisons, such as $\neq$.

\begin{example}  \label{ex:cause} \ \red{(ex. \ref{ex:cause00} cont.) \ We recall that the query}

\centerline{$\mc{Q}\!: \ \exists x \exists y ( \St(x) \land \R(x, y) \land \St(y))$}

\noindent is true in $D$, for which we want to identify causes.

Tuple ${\St(s_3)}$ is a counterfactual cause for ${\mc{Q}}$:
if ${\St(s_3)}$ is removed from ${D}$,
 ${\mc{Q}}$ is no longer true. So, it is an actual cause with empty contingency set; and its
responsibility is ${1}$. \  ${\R(s_4,s_3)}$ is an actual cause for ${\mc{Q}}$ with contingency set
${\{ \R(s_3,s_3)\}}$:
 if ${\R(s_4,s_3)}$ is removed from ${D}$, ${\mc{Q}}$ is still true, but further removing the contingent tuple ${\R(s_3,s_3)}$ makes ${\mc{Q}}$ false.
 The responsibility of ${\R(s_4,s_3)}$ is ${\frac{1}{2}}$.
 ${\R(s_3,s_3)}$ and ${\St(s_4)}$ are actual causes, with responsibility  ${\frac{1}{2}}$. \boxtheorem
\end{example}

\subsection{Database repairs}\label{sec:repairs} \vspace{-2mm}   We introduce the main ideas behind {\em database repairs} by means of an example. If only deletions and insertions of tuples are admissible updates, the ICs we consider in this work can be enforced only by deleting tuples from the database, not by inserting tuples (we consider repairs via updates of attribute-values in Section \ref{sec:arepairs}).

\vspace{-1mm}
\begin{example} \label{ex:rep} \  \red{Database $D = \{\Py(a), \Py(e), \Q(a,v), \Q(d,s), $\linebreak $\Rd(a,t), \Rd(d,v)\}$, shown in tabular form as}

\vspace{3mm}
\hspace*{0.3cm}
\begin{tabular}{c|c|}\hline
$\Py$&$P.1$\\ \hline
&$a$\\
&$e$\\ \hhline{~-}
\end{tabular}~~
\begin{tabular}{c|c|c|}\hline
$\Q$&$S.1$&$S.2$\\ \hline
& $a$ & $v$\\
& $d$ & $s$ \\ \hhline{~--}
\end{tabular}~~
\begin{tabular}{c|c|c|}\hline
$\Rd$&$R.1$&$R.2$\\ \hline
& $a$ & $t$\\
& $d$ & $v$\\
\hhline{~--}
\end{tabular}

\vspace{3mm}
\noindent is inconsistent with respect to (w.r.t.) the set of DCs $\Sigma = \{\kappa_1,\kappa_2\}$, with
  \begin{eqnarray}
  \kappa_1\!:&& \ \neg \exists x \exists y (\Py(x) \wedge \Q(x,y)), \label{eq:k1}\\
\kappa_2\!:&& \ \neg \exists x \exists y (\Py(x) \wedge \Rd(x,y)), \label{eq:k2}
\end{eqnarray}
which require that pricey products cannot be supplied or resold. \ That is,   \  $D \not \models \Sigma$; and we have to consider possible repairs for $D$.

A {\em subset-repair},  in short an {\em S-repair}, of $D$ w.r.t.  $\Sigma$ is a $\subseteq$-maximal subset of $D$ that is consistent w.r.t. $\Sigma$, i.e.  no proper superset is consistent. The following are
the S-repairs: \begin{eqnarray*}
D_1 &=& \{\Py(e), \Q(a,v), \Q(d,s), \Rd(a,t),\Rd(d,v)\},\\ D_2 &=& \{\Py(e), \Py(a), \Q(d,s), \Rd(d,v)\}.
\end{eqnarray*}
\ A {\em cardinality-repair},  in short a {\em C-repair}, of $D$ w.r.t. $\Sigma$ is a maximum-cardinality S-repair of $D$ w.r.t. $\Sigma$.  $D_1$  is
the only C-repair. \boxtheorem
\end{example}

For an instance $D$ and a set $\Sigma$ of DCs, the sets of S-repairs and C-repairs are denoted with $\nit{Srep}(D,\Sigma)$ and $\nit{Crep}(D,\Sigma)$, resp.

The definitions of S- and C-repairs can be generalized  to  sets $\Sigma$ of arbitrary ICs, for which both tuple deletions and insertions can be used as repair updates. This is the case, for example, for  inclusion dependencies. In these cases, repairs do not have to be subinstaces of the inconsistent instance at hand, $D$. Accordingly, one considers the symmetric difference, $D \Delta D' := (D \smallsetminus D') \cup (D' \smallsetminus D)$, between $D$ and a potential repair $D'$. On this basis, an S-repair  is an instance $D'$ that is  consistent with $\Sigma$, and makes $D \Delta D'$ minimal under set inclusion. Furthermore, $D'$ is a C-repair if it is an S-repair that also minimizes $|D \Delta D'|$.
\ Cf. \cite{Bertossi2011} for a survey of database repairs.

\subsection{Disjunctive answer-set programs}\label{sec:dasps}

We consider disjunctive Datalog programs $\Pi$ with stable model semantics \cite{eiterGottlob97}, a particular class of answer-set programs (ASPs) \cite{asp}. They consist of a set $E$ of ground atoms, called the {\em extensional database}, and a  finite number of rules of the form
\begin{equation}
A_1 \vee \ldots A_n \leftarrow P_1, \ldots, P_m, \n N_1, \ldots, \n N_k,\label{eq:rule0}
\end{equation}
with $0\leq n,m,k$, and the $A_i, P_j, N_s$ are positive atoms. The arguments in these atoms are constants or variables. The variables in the $A_i, N_s$ appear all among those
in the $P_j$.

The constants in program $\Pi$ form the (finite) Herbrand universe $U$ of the program. The {\em ground version} of
program $\Pi$, $\nit{gr}(\Pi)$, is obtained by instantiating the variables in $\Pi$ with all possible combinations of
values from $U$. The Herbrand base, $\nit{HB}(\Pi)$, of $\Pi$ consists of all the possible ground atoms obtained by instantiating the
predicates in $\Pi$ on $U$. A subset $M$ of $\nit{HB}(\Pi)$ is a (Herbrand) model of $\Pi$ if it contains $E$ and satisfies $\nit{gr}(\Pi)$, that is: For every
ground rule $A_1 \vee \ldots A_n \leftarrow P_1, \ldots, P_m, \n N_1, \ldots,
\n N_k$ of $\nit{gr}(\Pi)$, if $\{P_1, \ldots, P_m\} \subseteq M$ and $\{N_1, \ldots, N_k\} \cap M = \emptyset$, then
$\{A_1, \ldots, A_n\} \cap M \neq \emptyset$. $M$ is a {\em minimal model} of $\Pi$ if it is a model of $\Pi$, and no proper subset of $M$ is a model of $\Pi$. $\nit{MM}(\Pi)$ denotes the class of minimal models of $\Pi$. \ This definition applies in particular to positive programs, i.e. that do not contain negated atoms in rule bodies (i.e. the antecedents of the implications).

Now, take $S \subseteq \nit{HB}(\Pi)$, and transform $\nit{gr}(\Pi)$ into a new, positive program $\nit{gr}(\Pi)\!\downarrow \!S$ (i.e. without $\nit{not}$), as follows:
Delete every ground instantiation of a rule (\ref{eq:rule0})  for which $\{N_1, \ldots, N_k\} \cap S \neq \emptyset$. Next, transform each remaining ground instantiation of a rule (\ref{eq:rule0})  into $A_1 \vee \ldots A_n \leftarrow P_1, \ldots, P_m$. By definition, $S$ is a {\em stable model} of $\Pi$ iff $S \in \nit{MM}(\nit{gr}(\Pi)\!\downarrow \!S)$. A program $\Pi$ may have none, one or several stable models; and each stable model is a minimal model (but not necessarily the other way around) \cite{gelfond}.

Disjunctive answer-set programs have been used to specify database repairs \cite{barcelo,monica,Bertossi2011}. We will use them in Section \ref{sec:specTuples}.

\ignore{
+++

A disjunctive Datalog program  is stratified if its set of predicates ${\cal P}$ can be partitioned into
a sequence ${\cal P}_1, \ldots, {\cal P}_k$ in such a way that, for every  $P \in {\cal P}$:
\begin{enumerate}
\item If $P \in {\cal P}_i$ and predicate $Q$ appears in a head of a rule
with $P$, then $Q \in {\cal P}_i$.
\item If $P \in {\cal P}_i$ and $Q$ appears positively in the body of a rule that has $P$ in the head, then $Q
\in {\cal P}_{\!j}$, with $j \leq i$.
\item If $P \in {\cal P}_i$ and $Q$ appears negatively in the body of a rule that has $P$ in the head, then $Q
\in {\cal P}_{\!j}$, with $j < i$.
\end{enumerate}
If a program is stratified, then its stable models can be computed bottom-up by propagating data upwards from the
underlying extensional database, and making sure to minimize the selection of true atoms from the disjunctive heads.
Since the latter introduce a form of non-determinism, a program may have several stable models.
+++}

\section{ \ Causes and Database Repairs}\label{sec:cauRepair}

In this section we concentrate first on {\em tuple-based causes} as introduced in Section \ref{sec:tcause}, and establish a reduction to the tuple-based database repairs of Section \ref{sec:repairs}. In Section \ref{sec:abstract}, we provide an abstract definition of cause on the basis of an abstract repair-semantics.


\red{Before proceeding in more technical terms, it is worth giving a general idea about what we do next. First, it is well-known that checking an integrity constraint (IC) on a database can be done by posing an associated  query, or by defining a so-called violation view and checking its contents. The IC is satisfied as long as the query does not have an answer, i.e. it is false, or, equivalently, when the view has an empty contents. In this way, repairs of the database w.r.t. the IC can be put in correspondence with causes for query answers (or view contents): If there is a violation of the IC by a tuple (or a combination thereof), then there is a cause (with a contingency set) for a query answer (that exists due to the IC violation); and vice versa. \ In this direction, we  establish next the right correspondence between tuples that are left outside a repair, with some companion tuples (for participating in a violation of the IC) and tuples that are causes for the query answer, with some contingent companion tuples.}

 Now we show in precise terms how causes (represented by database tuples) for queries can be obtained from database repairs \cite{tocs}.
Consider the BCQ \begin{equation*}
\mc{Q}\!: \exists \bar{x}(P_1(\bar{x}_1) \wedge \cdots \wedge P_m(\bar{x}_m))
 \end{equation*}that is (possibly unexpectedly) true in  $D$: \ $D \models \mc{Q}$. Actual causes for $\mc{Q}$, their  contingency sets, and responsibilities can be obtained from database repairs. First,
$\neg \mc{Q}$ is logically equivalent to  the  denial constraint:
\begin{equation}
{{\kappa(\mc{Q})}\!: \ \ \neg \exists \bar{x}(P_1(\bar{x}_1) \wedge \cdots \wedge P_m(\bar{x}_m))}. \label{eq:qkappa}
\end{equation}

\noindent
So, if $\mc{Q}$ is true in $D$, \ $D$ is inconsistent w.r.t. $\kappa(\mc{Q})$, giving rise to repairs of $D$ w.r.t. $\kappa(\mc{Q})$.

Next, we build differences, containing a tuple $\tau$, between $D$ and  S-  or  C-repairs:  \begin{eqnarray}
 \mbox{(a) }  \nit{Diff}^s(D,\kappa(\mc{Q}), \tau)  &=&  \{ D \smallsetminus D'~|~ D' \in \nit{Srep}(D,\kappa(\mc{Q})),   \tau \in (D\smallsetminus D')\},~~~ \label{eq:s}\\
 \mbox{(b) }  \nit{Diff}^c(D,\kappa(\mc{Q}), \tau)  &=&  \{ D \smallsetminus D'~|~ D' \in \nit{Crep}(D,\kappa(\mc{Q})),  \tau \in (D\smallsetminus D')\}.~~~ \label{eq:c}
 \end{eqnarray}

\begin{proposition}\label{prop:tocs} \em \cite{tocs} \ For an instance $D$, a BCQ $\mc{Q}$, and its associated DC $\kappa(\mc{Q})$, it holds:
\begin{enumerate}[(a)]
\item $\tau \in D$ is an {actual cause} for $\mc{Q}$ \ iff \
$\nit{Diff}^s(D, \kappa(\mc{Q}), \tau) \not = \emptyset$. \item For each S-repair $D'$ with $(D\smallsetminus D') \in \nit{Diff}^s(D, \kappa(\mc{Q}), \tau)$,  \ $(D\smallsetminus (D' \cup \{\tau\}))$ is a subset-minimal contingency set for $\tau$. \item If { $\nit{Diff}^s(D$  $\kappa(\mc{Q}),  \tau) = \emptyset$}, then {$\rho(\tau)=0$}.
 Otherwise, { $\rho(\tau)=\frac{1}{|s|}$}, where {  $s \in \nit{Diff}^s(D,$ $\kappa(\mc{Q}), \tau)$} and there is no { $s' \in \nit{Diff}^s(D,\kappa(\mc{Q}), \tau)$} with { $|s'| < |s|$}.
\item $\tau \in D$ is a most responsible actual cause  for $\mc{Q}$ \ iff \
$\nit{Diff}^c\!(D,\kappa(\mc{Q}), \tau) \not = \emptyset$.
\end{enumerate} \vspace{-2mm}\boxtheorem
\end{proposition}

\begin{example} (ex. \ref{ex:cause} cont.) \label{ex:kappa} \  With the same instance $D$ and query $\mc{Q}$, we consider the
DC \
\begin{equation*}\red{\kappa(\mc{Q}):  \ \neg \exists x\exists y( \St(x)\wedge \R(x, y)\wedge \St(y))},
\end{equation*} which is not satisfied by $D$.

 Here, ${\nit{Srep}(D, \kappa(\mc{Q})) =\{D_1, D_2,D_3\}}$ and ${\nit{Crep}(D, \kappa(\mc{Q}))=\{D_1\}}$, with:
\begin{eqnarray*}
D_1&=& \{\R(s_4,s_3), \R(s_2,s_1), \R(s_3,s_3), \St(s_4), \St(s_2)\},\\
D_2 &=& \{ \R(s_2,s_1), \St(s_4),\St(s_2), \St(s_3)\},\\
  D_3 &=& \{\R(s_4,s_3), \R(s_2,s_1), \St(s_2),\St(s_3)\}.
  \end{eqnarray*}

For tuple \ $\R(s_4,s_3)$,  \ ${\nit{Diff}^s(D, \kappa(\mc{Q}), \R(s_4,s_3))=\{D \smallsetminus D_2\}}$ $= \{ \{ \R(s_4,s_3),$ $ \R(s_3,s_3)\} \} = \{\{ \R(s_4,s_3)\} \cup \Gamma_2\}$, with
$\Gamma_2 =$ \linebreak $ \{\R(s_3,s_3)\}$ as a minimum-cardinality contingency set for $\R(s_4,s_3)$, of size $1$. So,
 $\R(s_4,s_3)$ is an actual cause,  with responsibility $\frac{1}{2}$.

 Similarly, $\R(s_3,s_3)$ is an actual cause, with responsibility $\frac{1}{2}$.
\ For tuple $\St(s_3)$,  \  $\nit{Diff}^c(D, \kappa(\mc{Q}), \St(s_3)) =$ $ \{D \smallsetminus D_1\} =\{ \{\St(s_3)\} \}$. \
So, $\St(s_3)$
is an actual cause,  with responsibility $1$, i.e. a  {most responsible cause}.

Notice that $\R(s_4,s_3)$ is an actual cause whose minimum-cardinality contingency set $\Gamma_2$  is associated to an S-repair, $D_2$, that is not a C-repair; whereas $\St(s_3)$ is a maximum-responsibility actual cause whose
 minimum-cardinality contingency set, the empty set, is associated to the C-repair $D_1$. \boxtheorem
\end{example}

This connection between repairs and actual causes with their responsibilities can be extended to include actual causes for \red{{\em Unions of Boolean Conjunctive Queries} (UBCQs)} and repairs wrt. sets of DCs.

\begin{example}\label{ex:rc2cp} (ex. \ref{ex:rep} cont.) \ With the same database $D$,
consider the query (an UBCQs) $\mc{Q} := \mc{Q}_1 \vee \mc{Q}_2$, with
 \begin{eqnarray*}
 \mc{Q}_1\!:&& \exists x\exists y (\Py(x) \land \Q(x,y)),\\
 \mc{Q}_2\!:&& \exists x \exists y (\Py(x) \land \Rd(x,y)).
 \end{eqnarray*} It generates the set  of DCs: \ $\Sigma=\{\kappa_1,\kappa_2\}$, with
$\kappa_1$ and $\kappa_2$ as in (\ref{eq:k1}) and (\ref{eq:k2}), resp.  Here, $D \models \mc{Q}$ and, accordingly, $D$ is inconsistent w.r.t. $\Sigma$.

 The actual causes for $\mc{Q}$ in $D$ are: $\Py(a)$, $\Q(a,v)$, and $\Rd(a,t)$, with
$\Py(a)$ the most responsible cause. The only S-repairs for $D$ are:
 \begin{eqnarray*} D_1&=& \{ \Py(a),\Py(e), \Q(d,s),  \Rd(d,v)\},\\
D_2&=& \{\Py(e),\Q(a,v),\Q(d,s), \Rd(a,t),   \Rd(d,v)\};
\end{eqnarray*}  and $D_2$ is also the only C-repair.
\boxtheorem
\end{example}

It is also possible, the other way around, to characterize repairs in terms of causes and their contingency sets \cite{tocs}. Actually this latter connection can be used to obtain complexity results for
causality problems from repair-related computational problems \cite{tocs}. Most computational problems related to repairs, especially C-repairs, which are related to most responsible causes, are provably hard.
This is reflected in a high complexity for responsibility \cite{tocs} \ (cf. Section \ref{sec:disc} for some more details).

\subsection{Abstract causes from abstract repairs}\label{sec:abstract}

We can extrapolate  from the characterization of causes in terms of repairs that we have shown in this section, by starting
from an abstract {\em repair-semantics},  $\nit{Rep}^{\!{\sf S}\!}(D,\kappa(Q))$, which identifies a class of intended repairs of instance $D$ w.r.t. the DC $\kappa(Q)$. By definition, $\nit{Rep}^{{\sf S}\!}(D,\kappa(Q))$ contains instances for $D$'s schema that satisfy $\kappa(Q)$ and depart from $D$ in an ${\sf S}$-dependent minimal way \cite{Bertossi2011}. The most common repair semantics w.r.t. DCs is that of S-repairs, which are all subinstances of $D$. However, the minimality criterion does not have to be based on set inclusion (as is the case for S-repairs). Even more, the repairs do not have to be subinstances of $D$, even for DCs, as we will see in  Section \ref{sec:arepairs}.

\ More concretely, given a possibly inconsistent instance $D$, a general class of repair semantics can be characterized through an abstract partial-order relation, $\preceq_D$,\footnote{That is, satisfying reflexivity, transitivity and anti-symmetry, namely $D_1 \preceq_D D_2 \mbox{ and } D_2 \preceq_D D_1 \ \Rightarrow D_1 = D_2$.} on instances of $D$'s schema that is parameterized by $D$.\footnote{These general {\em prioritized repairs} based on this kind of priority relations were introduced in \cite{stawo}, where also different priority relations and the corresponding repairs were investigated.}
\ If we want to emphasize this dependence on the {\em priority relation} \ $\preceq_D$, we define the corresponding class of repairs of $D$ w.r.t. a set on ICs $\Sigma$ as:
  \begin{equation}\nit{Rep}^{{\sf S}^\preceq}(D,\Sigma) := \{D'~|~ D' \models \Sigma, \mbox{ and } D' \mbox{ is } \preceq_D\!\mbox{-minimal}\}.
  \end{equation}
This definition is general enough to capture different classes of repairs, and in relation to different kinds of ICs, e.g. those that delete old tuples and introduce new tuples to satisfy inclusion dependencies, and also repairs that change attribute values. In particular, it is easy to verify that  the classes of S- and C-repairs for DCs of Section \ref{sec:repairs} are particular cases of this definition.

If we assume that the repairs provided by the abstract repair semantics $\nit{Rep}^{{\sf S}\!}(D,\kappa(Q))$ are all sub-instances of $D$, and we let us inspire by (\ref{eq:s}), we can introduce:
\begin{equation}
\nit{Diff}^{\sf S}(D,\kappa(\mc{Q}), \tau) :=  \{ D \smallsetminus D'~|~ D' \in \reps, \  \tau \in (D\smallsetminus D')\}. \label{eq:srep}
\end{equation}

\begin{definition}\label{def:absCause} For an instance $D$, a BCQ $\mc{Q}$, and a class of repairs $\reps$:

\noindent (a)  $\tau \in D$ is an {\em actual {\sf S}-cause} for $\mc{Q}$  iff
$\nit{Diff}^{\sf S}(D, \kappa(\mc{Q}), \tau) \not = \emptyset$.

\noindent (b) For each  $D' \in \reps$ with $(D\smallsetminus D') \in \nit{Diff}^s(D, \kappa(\mc{Q}), \tau)$, $(D\smallsetminus (D' \cup \{\tau\}))$ is an \sfS-contingency set for $\tau$.

\noindent  (c) The \sfS-responsibility of an actual \sfS-cause is as in Section \ref{sec:tcause}, but considering only the cardinalities of  \sfS-contingency sets $\Gamma$. \boxtheorem
\end{definition}
It should be clear that actual causes as defined in Section \ref{sec:cauRepair} are obtained from this definition by using S-repairs. Furthermore, it is also easy to see that each  actual {\sf S}-cause accompanied by one of its {\sf S}-contingency sets falsifies query $\mc{Q}$ in $D$.

This abstract definition can be instantiated with different repair-semantics, which leads to different notions of cause.  It can also be modified in a natural way to define causes associated to repairs that may not be subinstances of the given instance. We will do this in the following subsection  by appealing to attribute-based repairs that change attribute values in tuples by \nn, a null value that is assumed to be a special constant in  the set $\mc{C}$ of constants in the data domain associated to the database schema. This will allow us, in particular,  to define causes at the attribute level (as opposed to tuple level) in a very natural manner.

A similar approach based on abstract repair semantics was taken in \cite{sum18,sum18corr} in order to introduce an abstract  {\em inconsistency measure} of a database w.r.t. a set on ICs.
\ \red{In Section \ref{sec:arepairs}, we instantiate the abstract semantics to define null-based causes from a particular but natural and practical notion of attribute-based repair.}

  \section{\red{Attribute-Based Causes}}\label{sec:arepairs}

\red{Causality in databases has been developed mostly in terms of {\em tuples} that are causes for query answers. However, this notion may suffer from a low level of granularity in that there could be certain {\em attribute values} in a tuple that may have more impact than others on a query answer. Defining attribute-level causes could be done directly. Instead, in this section, we appeal to the abstract notion of cause as related to an also abstract notion of repair, as we did in Section \ref{sec:abstract}. In order to do this, we appeal again to the connection between IC violation and the existence of answers to a query (c.f. beginning of Section \ref{sec:cauRepair}). However, instead of considering repairs of the database that are obtained by deletions of full tuples, we consider repairs that modify attribute values in tuples. }

\red{The most ``neutral", natural, and least arbitrary change on an attribute value one could attempt to eliminate a violation of an IC or a query answer (the idea behind actual causality) is the replacement of that value by a {\em null value}, in the spirit and the behavior  of {\footnotesize{\sf NULL}} in SQL databases. Accordingly, we first define repairs of databases in terms of changes of attribute values by a (unique and single) null value, and then we define causes at the attribute-value level, by identifying those attribute-values that are replaced by a null in a repair.}

\red{In order to identify attribute-values in tuples appearing in the original database, it is necessary to keep track of changes while having the capability to identify the original tuples. For this reason, we have to introduce
{\em unique, global and unchangeable identifiers} in database tuples. Furthermore, since the only admissible changes are by a null value, it is good enough to identify the (attribute) position where such a change takes place. Only the value in the original tuple is relevant in the end. For this reason, in the following we represent  values in tuples in the form $\nit{PredicateName}[\nit{tupleId}; \nit{position}]$, e.g. $R[{\sf t_4}; 3]$  for the tuple $R({\sf t_4}, a, b, c,d)$  refers to the value $c$ in the tuple (the identifier appears in the extra position $0$). Since the tuples ids are global, we could dispense with the predicate name if we wanted.}

  Database repairs that are based on changes of attribute values in tuples have been considered in \cite{Bertossi2011,tplp,IS08}, and implicitly in \cite{tkde} to hide sensitive information in a database $D$ via minimal virtual modifications of $D$. In the rest of this section, we make explicit this latter approach and exploit it to define and investigate attribute-based causality (cf. also \cite{tocs}). First we provide a motivating example.

\begin{example}  \label{ex:cause2}  Consider the database instance

\centerline{\red{$D = \{\St(s_2), \St(s_3), \R(s_3,s_1),\R(s_3,s_4),\R(s_3,s_5)\},$}}

\noindent and the query \
\begin{equation}
\red{\mc{Q}\!: \ \exists x \exists y (\St(x) \land \R(x, y)).} \label{eq:q}
\end{equation}  $D$ satisfies $\mc{Q}$, i.e. \ ${D \models \mc{Q}}$.

The three $\R$-tuples in $D$ are actual causes,
but clearly the  value $s_3$ for the first  attribute of $R$ is what matters in them, because it enables the join,   e.g. $D \models \St(s_3) \wedge \R(s_3,s_1)$. This is only indirectly captured through the occurrence of different values accompanying $s_3$ in the second attribute of $\R$-tuples as causes for $\mc{Q}$. \
Now, consider the instance
\begin{eqnarray*}
D_1 &=& \{\St(s_2), \St(s_3), \R(\nn,s_1),\R(\nn,s_4),\\ &&~~\R(\nn,s_5)\},
\end{eqnarray*}
where $\nn$ stands for the null value as used in SQL databases, which cannot be used to satisfy a join. \ Now, \ $D_1 \not \models \mc{Q}$. \ The same occurs with the instances $D_2 = \{\St(s_2), \St(\nn),$$ \R(s_3,s_1), \R(s_3,s_4),$ $ \R(s_3,s_5)\}$, \ and \ $D_3 = \{\St(s_2), \St(\nn),  \R(\nn,s_1),$ \linebreak $\R(\nn,s_4),$ $ \R(\nn,s_5)\}$, among others that are obtained from $D$ only through changes of attribute values by \nn.\footnote{Cf. also \cite[secs. 4, 5]{tplp} for an alternative repair-semantics based on both null- and tuple-based repairs w.r.t. general sets of ICs and their repair-programs. They could also be used to define a corresponding notion of cause.}
\boxtheorem
\end{example}

In the following we will assume the special constant $\nn$ may appear in database instances, and can be used to verify queries and constraints. We assume that all atoms with built-in comparisons, say $\nn \ \theta \ \nn$, and $\nn \ \theta \ c$, with $c$ a non-null constant, are all false for $\theta \in \{=,\neq, <, >, \ldots\}$. In particular, since a join, say  $R(\ldots, x) \wedge S(x,\ldots)$, can be written as $R(\ldots, x) \wedge S(x',\ldots) \wedge x=x'$, it can never be satisfied through \nn. This assumption is compatible with the use of {\footnotesize {\sf NULL}} in SQL databases (cf. \cite[sec. 4]{tplp} for a detailed discussion, also \cite[sec. 2]{tkde}). However, it should be clear that these basic assumptions on ``the logic" of \nn \ does not force us to bring SQL into our framework.

Consider an instance $D = \{\ldots, R(c_1, \ldots, c_n), \ldots\}$ that may be
inconsistent with respect to a set of DCs. The allowed repair updates are changes of
attribute
values by  \nit{null}, which is a natural choice, because this is a deterministic solution that appeals to {\em the} generic data value used in SQL databases to reflect the uncertainty and incompleteness in/of the database that inconsistency produces.\footnote{Repairs based on updates of attribute values using other constants of the domain have been considered in \cite{wijsen}. We think the developments in this section could be applied to them.}
As mentioned above, in order to keep track of changes, we introduce numerical
arguments in tuples as  global, unique tuple identifiers (tids), in position $0$. So, $D$ becomes $D = \{\ldots, R(i;c_1, \ldots, c_n),
\ldots\}$, with $i \in \mathbb{N}$.   With $\nit{id}(\tau)$ we denote the id of the tuple
$\tau \in D$, i.e.  $\nit{id}(R(i;c_1, \ldots, c_n)) = i$.

If  $D$ is updated to $D'$ by replacement of (non-tid) attribute values by \nn, and the value of the $j$-th attribute
in $R$, with $j>0$, is changed into \nn, then the change is captured as the string
$R[i;j]$, which identifies that the change was made in the tuple with id $i$ in the $j$-th {\em position} (or attribute) of predicate $R$.

More precisely, for a tuple $R(i;\bar{a})$, \ $R(i;\bar{a}){\tiny \frac{\nn}{j_1,\cdots, j_k}}$ denotes the the tuple that results from replacing the values in positions $j_i$ by \nn \ in $R(i;\bar{a})$. For example, for $R(8;a,b,a)$, \ $R(8;a,b,a)\frac{\nn}{2,3} = R(8;a,\nn,\nn)$.

These strings are collected
in the set:\footnote{The condition $s_i \neq \nn$ in its definition is needed in case the initially given instance already contain nulls.}
\ignore{\begin{eqnarray*}
\Delta^\nn(D,D') &:=& \{R[i;j]~|~ R(i;c_1, \ldots, c_j, \ldots, c_n) \in D, c_j \neq \nn, \mbox{ becomes}\\
&& \hspace*{4cm}R(i;c_1', \ldots,\nn, \ldots, c_n') \in D'\}.
\end{eqnarray*}}
\begin{eqnarray*}
\Delta^\nn(D,D') &:=& \{R[i;j]~|~ \exists R(i;\bar{a}) \in D, \mbox{ with } R(i;\bar{a}){\tiny \frac{\nn}{j_1,\cdots, j_k}} \in D', \\  && \hspace*{4.3cm}j \in \{j_1,\ldots, j_k\}, \mbox{ and } s_j \neq \nn\}.
\end{eqnarray*}

\noindent For example, if \ $D = \{R(1;a,b),
S(2;c,d), S(3;e,f)\}$ \ is changed into \ $D' =$ \linebreak $ \{R(1;a,\nn),$ $ S(2;\nn,d),$ $
S(3;\nn,\nn)\}$, then  $\Delta^\nn(D,D') = \{R[1;2],$ \linebreak $S[2;1],
S[3;1], S[3;2]\}$.

 The use of \nn \ is particularly useful to restore consistency w.r.t. DCs, which involve combinations of (unwanted) joins.

\begin{example}  \label{ex:cause3} (ex. \ref{ex:cause2} cont.) \ Still with the database instance
\begin{equation*}
\red{D = \{\St(s_2), \St(s_3), \R(s_3,s_1),\R(s_3,s_4),\R(s_3,s_5)\},}
\end{equation*}  consider the DC corresponding to the negation of query $\mc{Q}$ in (\ref{eq:q}): \
\begin{equation*}\kappa: \ \neg \exists x \exists y (\St(x) \land \R(x, z)).
\end{equation*} Since $D \not \models \kappa$, $D$ is inconsistent. \
The updated instance
\begin{equation*}D_2 = \{\St(s_2), \St(\nn), \R(s_3,s_1),\R(s_3,s_4),\R(s_3,s_5)\}
\end{equation*} is consistent (among others obtained by updates with \nn), i.e. \ $D_2 \models \kappa$.  \boxtheorem
\end{example}

\begin{definition}
A {\em null-based repair} of $D$ with respect to a set of DCs $\Sigma$ is a consistent
instance $D'$, such that $\Delta^\nn(D,D')$ is minimal under set
inclusion.\footnote{An alternative, but equivalent
formulation can be found in \cite{tkde}.}
$\nit{Rep}^\nit{null}(D,\Sigma)$ denotes the class of null-based repairs of
$D$ with respect to $\Sigma$.\footnote{Our setting allows for a uniform treatment of general and combined DCs, including those with (in)equality and other built-ins, FDs, and {\em key constraints} (KCs). However, for KCs in SQL databases, it is common  that {\scriptsize {\sf NULL}} is disallowed as a value for a key-attribute, among other issues. This prohibition, that we will ignore in this work, can be accommodated in our definition. For a detailed treatment of repairs w.r.t. sets of ICs that include FDs, see \cite[secs. 4,5]{tplp17}.}
\ignore{\footnote{We will exclude FDs from sets of DCs. There is no ``logical" problem in treating them uniformly with and as DCs, but FDs, and especially key constraints, are tricky in the presence of SQL's {\scriptsize {\sf NULL}}, e.g. a key-attribute is not allowed to take this value, among other issues. For a detailed treatment of repairs w.r.t. sets of ICs that include FDs, see \cite[secs. 4,5]{tplp}.}} \ A {\em cardinality-null-based repair} $D'$ minimizes $|\Delta^\nn(D,D')|$. \boxtheorem
\end{definition}
We can see that the null-based repairs are the minimal elements of the partial order between instances defined by: \ $D_1 \leq_D^\nn D_2$ \ iff \ $\Delta^\nn(D,D_1) \subseteq
\Delta^\nn(D,D_2)$.

\begin{example}\label{ex:nullReps} (ex. \ref{ex:kappa} cont.) \
Consider  instance $D$ with tuple ids now:
\begin{center}
\red{\begin{tabular}{r|c|c|} \hline
$\R$  & $R.1$ & $R.2$ \\\hline
1& $s_2$ & $s_1$\\
2& $s_3$ & $s_3$\\
3 & $s_4$ & $s_3$\\
 \hhline{~--}
\end{tabular} \hspace*{0.5cm}\begin{tabular}{r|c|c|}\hline
$\St$  & S.1  \\\hline
4& $s_2$ \\
5& $s_3$ \\
6 & $s_4$ \\
\hhline{~-}
\end{tabular}}
\end{center}
Equivalently, $D= \{\R(1;s_2,s_1), \R(2;s_3,s_3),\R(3;s_4,s_3),$\linebreak   $\St(4;s_2),$ $\St(5;s_3),$
$\St(6;s_4)\}$. It is inconsistent w.r.t. the DC: \\ \centerline{$\kappa\!: \ \neg
\exists x y (\St(x) \wedge \R(x, y) \wedge \St(y))$.}

Using just $R$ for $\R$, and $S$ for $\St$, to keep the presentation more compact,
the  class of null-based repairs, \
$\nit{Rep}^\nit{null}(D,\kappa)$, consists of:

\vspace{2mm}
$D_1= \{R(1;s_2,s_1),R(2;s_3,s_3), R(3;s_4,s_3),S(4;s_2), S(5;\nit{null})  ,
S(6;s_4)\}$,

$D_2 = \{ R(1;s_2,s_1), R(2;\nit{null},s_3), R(3;s_4,\nit{null}), S(4;s_2),
S(5;s_3),S(6;s_4)\}$,

$D_3 = \{ R(1;s_2,s_1), R(2;\nit{null},s_3), R(3;s_4,s_3), S(4;s_2),
S(5;s_3),S(6;\nn)\}$,

$D_4 = \{ R(1;s_2,s_1), R(2;s_3,\nit{null}), R(3;s_4,\nit{null}), S(4;s_2),
S(5;s_3),S(6;s_4)\}$,

$D_5 = \{ R(1;s_2,s_1), R(2;s_3,\nit{null}), R(3;\nn,s_3), S(4;s_2),
S(5;s_3),S(6;s_4)\}$,

$D_6 = \{ R(1;s_2,s_1), R(2;s_3,\nit{null}), R(3;s_4,s_3), S(4;s_2),
S(5;s_3),S(6;\nn)\}$.

\vspace{2mm}
\noindent Then,  $\Delta^\nn(D,D_2) = \{ \R[2;1], \R[3;2]\}$, \ $\Delta^\nn(D,D_3) =$ \linebreak $ \{\R[2;1],$ $ \St[6;1]\}$ and $\Delta^\nn(D,D_1) = \{\St[5;1]\}$. The latter is a cardinality-null-based repair. \boxtheorem
\end{example}

According to the motivation provided at the beginning of this section, and drawing inspiration from the generic construction in (\ref{eq:srep}),
we can now define causes using as a concrete repair semantics the class of null-based repairs of $D$. Since
repair actions in this case are attribute-value changes,  causes can be defined at both the tuple- and attribute-levels; and the same applies to the definition of responsibility. In order to do both,  we first need to refine
(\ref{eq:srep}), which captures how repairs differ from the original instance.

\begin{definition} Consider an instance $D$, a tuple $\tau\!: \ R(i;c_1,\ldots,c_n) \in D$, a BCQ $\mc{Q}$, and the associated DC $\kappa(\mc{Q})$: \
(a) $R[i;j]^D$ denotes the attribute value that appears in the $j$th position of the $R$-tuple in $D$ with tid $i$. \ (b) For an attribute value $\nu = R[i;j]^D$:
\begin{eqnarray}
\nit{Diff}^\nn(D,\kappa(\mc{Q}), \nu) &:=&  \{ \Delta^\nn(D,D')~|~ D' \in \nit{Rep}^\nn(D,\kappa(\mc{Q})),\label{eq:ndiff}\\ &&\hspace*{2.5cm}  R[i;j] \in \Delta^\nn(D,D')\}. \nonumber
\end{eqnarray}

\vspace{-10mm}\phantom{oo} \boxtheorem
\end{definition}
Notice that (\ref{eq:ndiff}) is not  a particular case of  (\ref{eq:srep}), because it does not contain full tuples. \
In (a) we have ``positioned" attribute values\ignore{, and they  will be usually denoted with $t$}. For example,
for $R(4; a, c, a) \in D$,  $R[4;1]^D = a$,  $R[4;2]^D = c$ and $R[4;3]^D = a$. So, there are two different attribute values $a$, because they appear in different positions.

\begin{definition} \label{def:attCuases}
 For  instance $D$, a   BCQ
$\mc{Q}$, and $\nu = R[i;j]^D$:
   \begin{enumerate}
 \item [(a)] Value \ $\nu$ is an {\em attribute-null-based (actual) cause} for $\mc{Q}$ \  iff $\nit{Diff}^\nn(D,$ $\kappa(\mc{Q}),\nu)$ $ \neq \emptyset$,
i.e. $\nu$  is a cause if it is changed into a null in some repair. \ When there is $D' \in \nit{Rep}^\nn(D,\kappa(\mc{Q}))$ with $\Delta^\nn(D,D') = \{R[i;j]\}$, $\nu$ is called a {\em counterfactual} attribute-null-based cause.

\item [(b)] Tuple $\tau \in D$, with $i' = \nit{id}(\tau)$, is a \ {\em tuple-null-based (actual) cause} for $\mc{Q}$ \  if  some $S[i';j']^D$ is an {\em attribute-null-based cause} for $\mc{Q}$, i.e. the whole tuple $\tau$ is a cause if at least one of its attribute values is changed into a null in some repair.

\item [(c)] The responsibility, $\rho^{\mbox{\small \it a-null}}(\nu)$, of  an  attribute-null-based cause $\nu$ for $\mc{Q}$, is the inverse of \
$\nit{min}\{|\Delta^\nn(D,D')|~:~R[i;j] \in  \Delta^\nn(D,$ $D'),$  $\mbox{and } D' \in  \nit{Rep}^\nn(D,\kappa(\mc{Q}))\}$. Otherwise, if $\nu$ is not an attribute-null-based cause, its responsibility is $0$.

\item [(d)]
The responsibility, $\rho^{\mbox{\small \it t-null}}(\tau)$, of  a tuple-null-based cause $\tau$ for $\mc{Q}$, is the inverse of \
$\nit{min}\{|\Delta^\nn(D,D')|~:~ i' = \nit{id}(\tau), \ S[i';j'] \in  \Delta^\nn(D,D'), \mbox{ for some}$ $ S, j'$, $ \mbox{ and}$ $ D' \in$  $  \nit{Rep}^\nn(D,$ $\kappa(\mc{Q}))\}$. Otherwise, if $\tau$ is not a tuple-null-based cause, its responsibility is $0$.
\boxtheorem
 \end{enumerate}
\end{definition}

In cases (c) and (d) we minimize over the number of changes in a repair. However, in case (d), of a tuple-cause,
any change made in one of its attributes is considered in the minimization. For this reason, the minimum may be smaller than the one for a fixed attribute
value change; and so the responsibility at the tuple level may be greater than that at the attribute level. More precisely,
 if $\tau = R(i;c_1, \ldots, c_n) \in D$, and $c_j =R[i;j]^D$ is an attribute-null-based cause, then: \
$\rho^{\mbox{\small \it a-null}}(R[i;j]^D) \leq \rho^{\mbox{\small \it t-null}}(\tau)$.

\begin{example} (ex. \ref{ex:nullReps} cont.) \ Consider $\R(2;s_3,s_3) \in D$. Its projection on its first (non-id) attribute, $s_3=\R[2;1]^D$, is an attribute-null-based
cause since $\R[2;1] \in
\Delta^\nn(D,D_2)$. Also $\R[2;1] \in
\Delta^\nn(D,D_3)$.

 Since $|\Delta^\nn(D,D_2)| = |\Delta^\nn(D,D_3)| = 2$, \ it holds \ $\rho^{\mbox{\small \it a-null}}(\R[2;1]^D)$ $= \frac{1}{2}$. \ Furthermore,
$\R(2;s_3,s_3)$ is  a tuple-null-based cause for $\mc{Q}$, with \linebreak $\rho^{\mbox{\small \it t-null}}(\R(2;s_3,s_3)) = \frac{1}{2}$.
\boxtheorem
\end{example}

\begin{example} (ex. \ref{ex:cause3} cont.)\label{ex:cause4} \ The instance with tids is $D= \{\St(1;s_2),$ \linebreak $ \St(2;s_3),$ $ \R(3;s_3,s_1),$ $ \R(4;s_3,s_4),$ $ \R(5;s_3,s_5)\}$. The \linebreak null-based repairs are $D_1$ and $D_2$, with  $\Delta^\nn(D,D_1) = \{\R[3;1],$ \linebreak $\R[4;1], \R[5;1]\}$ and  $\Delta^\nn(D,D_2) = \{\St[2;1]\}$.

The values $s_3=\R[3;1]^D, s_3=\R[4;1]^D, s_3 = \R[5;1]^D,$ \linebreak $ s_3 = \St[2;1]^D$ are all attribute-null-based causes for $\mc{Q}$.

Notice that \ $\rho^{\mbox{\small \it a-null}}(\R[3;1]^D) =  \rho^{\mbox{\small \it a-null}}(\R[4;1]^D) = $ \linebreak $\rho^{\mbox{\small \it a-null}}(\R[5;1]^D) = \frac{1}{3}$, while $\rho^{\mbox{\small \it a-null}}(\R[3;2]^D) =  $ \linebreak $\rho^{\mbox{\small \it a-null}}(\R[4;2]^D) = \rho^{\mbox{\small \it a-null}}(\R[5;2]^D) = 0$, and that the value ($s_3$) in the first argument of the $\R$-tuples has a non-zero responsibility, while the values in the second argument have responsibility $0$. \boxtheorem
\end{example}

Notice that the definition of tuple-level responsibility, i.e. case (d) in Definition \ref{def:attCuases}, does not take into account that a same id, $i$, may appear
several times in a $\Delta^\nn(D,D')$. In order to do so, we could redefine the size of the latter by taking into account those multiplicities. For example, if we decrease the size of the $\Delta$ by one
with every repetition of the id, the
responsibility  for
a cause may (only) increase, which makes sense.

It is not difficult to define attribute-based causality in direct counterfactual terms as in Section \ref{sec:tcause}. Such a characterization is implicity given in Proposition \ref{prop:updates} below, for which we need some notation.

\begin{definition} \label{def:update}  Consider a database $D$ with unique tids. \ (a) An {attribute-null-based update} for $D$ is a set $U$ with elements of the form $R[i;j]$, where $R$ is a relational predicate of arity $n+1$, $i$ is the tid appearing in $D$ in a tuple's  $0$th-position, and $j \in \{1,\ldots,n\}$. The result applying $U$ to $D$ is the database instance $U\!\circ \! D$ defined by:
\begin{eqnarray*}
U\!\circ \! D &:=& \{\tau \in D~|~ id(\tau) \mbox{ does not appear in } U\} \cup\\ &&\hspace*{0.5cm}\{R(i;\bar{a}){\tiny \frac{\nn}{j_1,\cdots, j_k}}~|~ R(i;\bar{a}) \in D \mbox{ and } R[i;j_1], \ldots, R[i;j_k] \in U\}. \hspace{2.3mm}  \Box
\end{eqnarray*}
\end{definition}

That is, $U\circ D$ is obtained from $D$ and $U$ by keeping all tuples of $D$ whose tids do not appear in $U$, and the other tuples in $D$ are kept, but when $R[i;j] \in U$, the attribute value in position $j$ of tuple with tid $i$ is replaced by $\nit{null}$.

\begin{example} \ \label{ex:last} (ex. \ref{ex:cause4} cont.) \ For the instance $D= \{\St(1;s_2), \St(2;s_3),$ \linebreak $ \R(3;s_3,s_1),$ $ \R(4;s_3,s_4),$ $ \R(5;s_3,s_5)\}$ \ and \ $U =$ \linebreak $ \{\R[3;1], \R[4;1],$ $ \R[5;1]\}$, \ $U \circ D = \{\St(1;s_2), \St(2;s_3),$  $\R(3;\nn,s_1),$  $ \R(4;\nn,s_4),$ $ \R(5;\nn,s_5)\}$. \boxtheorem
\end{example}


\begin{proposition} \label{prop:updates} \em For a database $D$ with tids, a BCQ $\mc{Q}$, and an attribute value $\nu = R[i;j]^D$: \ (a) $\nu$  is a counterfactual attribute-null-based  cause for $\mc{Q}$ in $D$\ iff \ for the  update $U =\{R[i;j]\}$ for $D$: \  $D \models \mc{Q}$, but $U \circ D \not \models \mc{Q}$. \ (b) $\nu$ is an actual attribute-null-based cause for $\mc{Q}$ \ iff \ there is an update $U$ of $D$ with $R[i;j] \notin U$, such that
$\nu$ is a counterfactual attribute-null-based  cause for $\mc{Q}$ in $U \circ D$. \boxtheorem
\end{proposition}

In Section \ref{sec:abc} we will provide repair-programs for null-based repairs, which can be used as a basis for specifying and computing attribute-null-based causes.

\section{ \ Specifying Tuple-Based Causes} \label{sec:specTuples} \vspace{-2mm} Given a database $D$, it is possible to specify the S-repairs of $D$ w.r.t. a set $\Sigma$ of DCs by means of an ASP   $\Pi(D,\Sigma)$, in the sense that the set, $\nit{Mod}(\Pi(D,\Sigma))$, of its stable models is  in one-to-one correspondence with  $\nit{Srep}(D,\Sigma)$ \cite{monica,barcelo} (cf. \cite{Bertossi2011} for more references).  In this section we will show that these {\em repair-programs} can be used as a basis for causality-related specifications and computations.
However, to ease the presentation, we
consider a single denial constraint\footnote{It is possible to consider combinations of DCs and FDs, corresponding to UCQs, possibly with $\neq$, by appealing to the extensions presented in \ \cite{tocs}.}
\begin{equation}\kappa\!:  \neg \exists \bar{x}(P_1(\bar{x}_1)\wedge \dots \wedge P_m(\bar{x}_m)). \label{eq:4lp}\end{equation}

Although necessary for attribute-based repairs but not for S-repairs, it is useful on the causality side to assume, as we did in Section \ref{sec:arepairs}, that tuples have unique, global tuple identifiers (tids), i.e. every tuple $R(\bar{c})$ in $D$ is represented as $R(t;\bar{c})$ for some integer $t$ in an extra, $0$th position, that is not used by any other tuple in $D$. Furthermore, for the repair-program we introduce a nickname predicate $R'$ for every predicate $R \in \mc{R}$ that has (yet) an extra, last positioned attribute to hold an annotation from the set $\{\sf{d}, \sf{s}\}$, for ``delete" and ``stays", resp. \ Nickname predicates are used to represent and compute repairs.

The repair-program, $\Pi(D,\{\kappa\})$, for $D$ and $\kappa$ contains all the tuples in $D$ as facts (with tids), plus the following rules:
\begin{eqnarray}
P_1'(t_1;\bar{x}_1,\sfd)\vee \cdots \vee P_m'(t_n;\bar{x}_m,\sfd) &\leftarrow& P_1(t_1;\bar{x}_1), \dots, P_m(t_m;\bar{x}_m). \label{eq:rule}\\
P_i'(t_i;\bar{x}_i,\sfs) &\leftarrow& P_i(t_i;\bar{x}_i), \ \nit{not} \ P_i'(t_i;\bar{x}_i,\sfd), \ i=1,\cdots,m, \nonumber
\end{eqnarray}
 where the $t_i, \bar{x}_i$ are all variables. \ Here, the first rule captures in its body a violation of DC $\kappa$, see  (\ref{eq:4lp}); and the head, i.e. the consequent, offers all the alternative tuples deletions that can solve that violation. \ The second (set of) rule(s) basically captures inertia: the repairs keep the original tuples that have not been deleted.

A stable   model $M$ of the program determines a repair $D'$ of $D$: \ $D' := \{P(t;\bar{c})~|$  $P'(t;\bar{c},\sfs) \in M\}$, and every repair can be obtained in this way \cite{monica}. The semantics of stable model semantics ensures that only a minimal set of tuples are deleted.

For an FD, say $\varphi\!: \ \neg \exists xyz_1z_2vw(R(x,y,z_1,v) \wedge R(x,y,z_2,w) \wedge z_1 \neq z_2)$, which makes the third attribute functionally depend upon the first two, the repair-program contains the rules:
\begin{eqnarray*}
R'(t_1;x,y,z_1,v,\sfd) \vee R'(t_2;x,y,z_2,w,\sfd) &\leftarrow& R(t_1;x,y,z_1,v), R(t_2;x,y,z_2,w),\\ && \hspace*{4.1cm} z_1 \neq z_2.\\
\!\!\!R'(t;x,y,z,v,\sfs) &\leftarrow& R(t;x,y,z,v),  \nit{not} \ R'(t;x,y,z,v,\sfd).
\end{eqnarray*}
 For sets of DCs and FDs, the repair-program can be made non-disjunctive by moving all the disjuncts but one, in turns, in negated form to the body of the rule \cite{monica,barcelo}. For example, the rule
$P(a) \vee R(b) \leftarrow \nit{Body}$, can be written as the two rules \ $P(a)  \leftarrow \nit{Body}, \nit{not} R(b)$ and $R(b) \leftarrow \nit{Body}, \nit{not} P(a)$. Still the resulting program may be {\em non-stratified}
if there is recursion via negation \cite{gelfond}, as in the case of FDs, and  DCs with self-joins.

\begin{example} (ex. \ref{ex:kappa} cont.) \label{ex:kappa2} \ For the DC

\centerline{$\kappa(\mc{Q})$:  \ $\neg \exists x\exists y( \St(x)\wedge \R(x, y)\wedge \St(y))$,}

\noindent  the repair-program contains the facts (with tids): \
$\R(1;s_4,s_3)$, \linebreak $\R(2;s_2,s_1)$, $\R(3;s_3,s_3)$, $ \St(4;s_4)$,  $\St(5;s_2)$, $\St(6;s_3)$;\linebreak and the rules: \vspace{-1mm}
\begin{eqnarray}
\St'(t_1;x,\sfd) &\vee& \R'(t_2;x,y,\sfd) \vee \St'(t_3;y,\sfd) \leftarrow \nonumber\\
&& ~~~~~\St(t_1;x), \R(t_2;x, y),\St(t_3;y). \label{eq:repRule}\\
\St'(t;x,\sfs) &\leftarrow& \St(t;x), \ \nit{not} \ \St'(t;x,\sfd). \ \ \ \ \mbox{ etc. } \nonumber
\end{eqnarray}
This repair-program has three stable models, $M_1, M_2, M_3$; with repair $D_1$ corresponding to the stable model
\begin{eqnarray*}
M_1 = \{\R'(1;s_4,s_3,\sfs), \R'(2;s_2,s_1,\sfs), \R'(3;s_3,s_3,\sfs),\\ ~~~~~~~~~~ \St'(4;s_4,\sfs), \St'(5;s_2,\sfs),
\St'(6;s_3,\sfd)\} \cup D. \end{eqnarray*}
The repair $D_1$ is read off from $M_1$ by keeping only tuples annotated with \sfs. \boxtheorem
\end{example}

If there are more that one DC or FD, one can build the repair-program exactly as above by introducing one rule of the form (\ref{eq:rule}) per DC or FD (cf. Example \ref{ex:allNew}).

Now, in order to specify causes for a CQ $\mc{Q}$ by means of a repair-program, we move on to the associated DC $\kappa(\mc{Q})$, and
 we concentrate, according to  (\ref{eq:s}), on the differences between $D$ and its repairs w.r.t. $\kappa(\mc{Q})$. These differences become represented by
$\{P(t;\bar{c})~|~P(t;\bar{c},\sfd)$ $ \in M\}$, the deleted tuples, with $M$ a stable model of the repair-program. They are used to compute actual causes and their $\subseteq$-minimal contingency sets, both expressed in terms of  tids.

 The actual causes for the query can be represented by their tids, and can be obtained by posing simple queries to the program under the {\em uncertain or brave} semantics that makes true what is true in {\em some} model of the repair-program.\footnote{As opposed to the {\em skeptical or cautious} semantics that sanctions as true what is true in {\em all} models. Both semantics as supported by the DLV system \cite{dlv}.} That is, the query is about the values for tids $t$ that are consequences of the program under the brave semantics, i.e. for which $\Pi(D,\{\kappa(\mc{Q})\}) \models_\nit{brave} \nit{Cause}(t)$ holds.  Here, the \nit{Cause} predicate has to be defined on top of $\Pi(D,$ $\{\kappa(\mc{Q})\})$. In the case of Example \ref{ex:kappa2},  by the rules:
 \begin{eqnarray}
 \nit{Cause}(t) &\leftarrow& R'(t;x,y,\sfd) \label{eq:cause1}\\
 \nit{Cause}(t) &\leftarrow& S'(t;x,\sfd).\label{eq:cause2}
 \end{eqnarray}

In order to represent contingency sets for a cause, given the repair-program for a DC $\kappa(\mc{Q})$, a new binary predicate $\nit{CauCont}(\cdot,\cdot)$ is introduced, which will contain a tid for cause in its first argument, and a tid for a tuple belonging to its contingency set. Intuitively, $\nit{CauCont}(t,t')$ says that $t$ is an actual cause, and $t'$ accompanies  $t$ as a member of the former's contingency set (as captured by the repair at hand or, equivalently, by the corresponding stable model).
More precisely, for each pair of  predicates $P_i, P_j$ in $\kappa(\mc{Q})$ (not necessarily different; they could be the same if the query has self-joins or we have a union of CQs, and then, several DCs), introduce the rule
$$\nit{CauCont}(t,t') \leftarrow P_i'(t;\bar{x}_i,\sfd),  P_j'(t';\bar{x}_j,\sfd), t\neq t',$$ with the inequality condition only when $P_i$ and $P_j$ are the same predicate (it is superfluous  otherwise).

\begin{example} (ex.  \ref{ex:kappa2} cont.) \label{ex:causes} \ In order to compute causes with contingency sets, the repair-program  can be extended with the following rules:
\begin{eqnarray*}
\nit{CauCont}(t,t') &\leftarrow& \St'(t;x,\sfd), \R'(t';u,v,\sfd).\\
\nit{CauCont}(t,t') &\leftarrow& \St'(t;x,\sfd), \St'(t';u,\sfd), t\!\neq\! t'.\\
\nit{CauCont}(t,t') &\leftarrow& \R'(t;x,y,\sfd), \St'(t';u,\sfd). \\
\nit{CauCont}(t,t') &\leftarrow& \R'(t;x,y,\sfd), \R'(t';u,v,\sfd), t\neq t'.
\end{eqnarray*}
For the stable model $M_2$ corresponding to repair $D_2$, we obtain $\nit{CauCont}(1,3)$ and $\nit{CauCont}(3,1)$, from the repair difference $D \smallsetminus D_2 = \{\R(s_4,s_3),$ $\R(s_3,s_3)\}$. \boxtheorem
\end{example}

We can use extensions of ASP with set- and  numerical aggregation to build the contingency set associated to a cause, e.g. the DLV system \cite{dlv}  by means of its DLV-Complex extension \cite{calimeri08,calimeri09} that supports set membership and union as built-ins. We introduce a binary predicate $\nit{preCont}$ to hold a cause (id) and a possibly non-maximal set of elements from its contingency set, and the following rules:
\begin{eqnarray}
\nit{Cont}(t, \{\}) &\leftarrow& \nit{Cause}(t), \ \nit{not} \ \nit{Aux}_C(t) \label{eq:cont3}\\
 \nit{Aux}_C(t) &\leftarrow& \nit{CauCont}(t,t') \label{eq:cont4}\\
\nit{preCont}(t,\{t'\}) &\leftarrow& \nit{CauCont}(t,t'). \label{eq:cont1}\\
\nit{preCont}(t, \nit{\#union}(C,\{t''\})) &\leftarrow& \nit{CauCont}(t,t''), \nit{preCont}(t, C), \label{eq:cont2}\\
 &&\nit{not} \ \nit{\#member}(t'',C).\nonumber  \\
 \nit{Cont}(t, C) &\leftarrow& \nit{preCont}(t, C), \ \nit{not} \ \nit{HoleIn}(t,C). \label{eq:cont5}\ignore{\ \ \ \ \ \ \ \ \ \ \mbox{\small (maximal sets)}}\\
\nit{HoleIn}(t,C) &\leftarrow& \nit{preCont}(t, C), \nit{CauCont}(t,t'), \label{eq:cont6} \\ && \nit{not} \ \nit{\#member}(t',C). \nonumber
\end{eqnarray}

\ignore{\comlb{Do we need the last two rules? Wouldn't the first two do the job? I think we need them, because we would -otherwise- produce many non-maximal contingency atoms.}
\comlb{The cases above do not cover the empty set as contingency set.} }

\noindent Rules (\ref{eq:cont3}) and (\ref{eq:cont4}) associate the empty contingency set to counterfactual causes, i.e. those that do not have to be accompanied by any other tuple to become an actual cause. For a non-counterfactual cause, rules (\ref{eq:cont1}) and (\ref{eq:cont2}) build its, possibly non-maximal, ``contingency sets" (actually, subsets of contingency sets) within a repair or stable model by starting from a singleton, and adding additional elements from the contingency set. Rules (\ref{eq:cont5}) and (\ref{eq:cont6}), which use the auxiliary predicate $\nit{HoleIn}$, make sure that a set-maximal contingency set is built from a pre-contingency set to which nothing can be added.

The responsibility for an actual  cause $\tau$, with tid $t$, and associated to a repair $D'$ (i.e. with $\tau \notin D'$), and corresponding to a model $M$ of the extended repair-program, can be computed by counting the number of tids $t'$ for which
$\nit{CauCont}(t,t') \in M$. This responsibility will be maximum within a repair (or model): \
$\rho(t,M) := 1/(1+ |d(t,M)|)$, where $d(t,M) := \{\nit{CauCont}(t,t') \in M\}$. This value can be computed by means of the {\em count} function, supported by  DLV  \cite{aggreg},   as follows: \
\begin{equation}
\mbox{\nit{pre-rho}}(t,n+1) \leftarrow  \#\nit{count}\{t' : \nit{CauCont}(t,t')\} = n. \label{eq:prho}
\end{equation}
 \noindent The local responsibility for a cause $t$ in a given model $M$ is then computed by: \
 \begin{equation}
\rho^M(t) := \frac{1}{m}, \label{eq:ro}
\end{equation}
with $\mbox{\nit{pre-rho}}(t,m) \in M$,
 or, equivalently, as \ $1/|d(M)|$, with \ $d(M) := \{$ $P(t';\bar{c},\sfd)~|$ $P(t';\bar{c},\sfd) \in M\}$.

{\em The repair-program extended with all the rules added for causality specification and computation will be called the ``causality-program"}. We will keep denoting with $\Pi$ the extension of the repair-program. \
Each model $M$ of the causality program, due to its correspondence with an S-repair, will return, for a given tid $t$ that is an actual cause, a minimum-cardinality contingency set $\Gamma_M(t)$ for $t$ {\em within that model}: no proper subset is a contingency set for $t$ in that model. However, $M$ may not correspond to a C-repair, and $\Gamma_M(t)$ (or its cardinality) may not make $t$ a maximum-responsibility actual cause (cf. Proposition \ref{prop:tocs} and Example \ref{ex:kappa}).

Actually, if we want maximum-responsibility causes,  we look for tids $t$ for which $\rho(t) := \mbox{max}\{\rho^M(t)~|~M \mbox{ is a model of } \Pi \mbox{ with } \nit{Cause}(t) \in M\}$, which would be, in principle, an off-line computation, i.e. not within the program.

However, if we are interested only in maximum-responsibility actual causes, we have an alternative way to proceed, fully within the causality program, as we now describe. We know from Proposition \ref{prop:tocs} that maximum-responsibility causes are associated to
C-repairs, and any C-repair will have a global maximum-cardinality from which the global maximum-responsibility for a cause can be obtained. Accordingly, it is good enough to specify and compute only C-repairs.

C-repairs can be specified by adding {\em weak-program-constraints} (WPCs) \cite{buca,dlv} to the  repair-programs we introduced above \cite{tplp}. In this case, since we want repairs that minimize the number of deleted tuples, for each database predicate $P$, we introduce the WPCs into the causality program:$$:\sim \  P'_i(t;\bar{x}_i,\sfd)., \ \ \ i=1, \ldots, m.$$

\noindent In a model $M$ of the so-extended causality program, the body can be satisfied, and then the program constraint violated, but the {\em number} of violations is kept to a minimum (among the models of the program without the WPCs). A causality-program with these WPCs specifies repairs that minimize the number of deleted tuples. As a consequence, {\em minimum-cardinality} contingency sets and maximum responsibilities can be computed (and only them), as above, but with the extended causality-programs.\footnote{In contrast, {\em hard} program-constraints, of the form \ $\leftarrow \nit{Body}$, eliminate the models where they are violated, i.e. where \nit{Body} is satisfied. WPCs as those above are sometimes denoted with \ $\Leftarrow \  P'(t;\bar{x},\sfd)$. If we used the hard program constraint \ $\leftarrow \  P'(t;\bar{x},\sfd)$ instead of the WPCs, we would be prohibiting tuple deletions. This would result in the empty set of models or just the original $D$ in case the latter is consistent. }

\begin{example}\label{ex:wcs} (ex. \ref{ex:causes} cont.) \ If we add to the causality-program $\Pi$ the WPCs:
\begin{eqnarray*}
&:\sim& \  \R'(t;x,y,\sfd),  \ \mbox{ and}\\
&:\sim& \  \St'(t;x,\sfd),
 \end{eqnarray*}the only model will be $M_1$, from which the maximum-responsibility actual cause $\St(6;s_3)$ can be obtained,  with its maximum-responsibility: \linebreak $\rho(\St(6;s_3))$  $ = \frac{1}{d(M_1)}, \mbox{ with } d(M_1) = |\{\St'(6;s_3,\sfd)\}| =1$, which, in its turn, can be obtained by means of the query: \ \
$:\!\!- \ \mbox{\nit{pre-rho}}(6,n)$, under the brave semantics, getting the value $n=1$. \
The other  non-maximum responsibility actual causes in Example \ref{ex:kappa2} are not obtained, because they are associated to the non-maximum-cardinality repair $D_2$ (or model $M_2$). \boxtheorem
\end{example}

More generally, if we are interested in the responsibility of possibly non-maximum-responsibility actual causes, (with tid) ${\sf t}$, we can get rid of the WPCs above, and, as above, pose to the causality-program the query: \  $:\!\!- \ \mbox{\nit{pre-rho}}({\sf t},n)$ under the brave semantics, with $n$ a variable. The result will be of the form: \ $\mbox{\nit{pre-rho}}({\sf t},{\sf n}_1), \ldots, \mbox{\nit{pre-rho}}({\sf t},{\sf n}_k)$, where $k$ is the number of S-repairs where ${\sf t}$ is deleted, and the ${\sf n}_i$ are the different values taken by variable $n$ in them. The minimum of the ${\sf n}_i$ is used to compute ${\sf t}$'s responsibility.\ignore{\footnote{Equivalently, the query: \ $:\!\!- \ \nit{rho}({\sf t},m)$ can be posed, with variable $m$; and the maximum answer value for $m$ is chosen.}}

\ignore{
\red{
\begin{example}\label{ex:all} Consider instance $D = \{R(1;a,b), R(2;a,c), S(3;b,c), S(4;b,d)\}$ and the DCs: \ $\kappa_1\!: \ \neg \exists x y z(R(x,y) \wedge R(x,z) \wedge y \neq z)$, and \ $\kappa_2\!: \ \neg \exists x y z v(R(x,y) \wedge S(y,z) \wedge S(y,v) \wedge   v \neq z)$. The repair-program has three stable models, $M_1, M_2, M_3$, corresponding to the S-repairs: \ $D_1 = \{R(2;a,c), S(3;b,C), S(4;b,d)\}$, \ $D_2 = \{R(1;a,b),$ $ S(4;b,d)\}$, \ $D_3 = \{R(1;a,b),  S(3;b,c)\}$. Only the first is a C-repair. For the tuple with tid $2$ the pre-responsibility is obtained with the query: \  $:\!\!- \ \mbox{\nit{pre-rho}}(2,n)$, which under the brave semantics gives  the numbers of tuples deleted together with tuple $4$ in S-repairs. We obtain $1$ and $1$ from $M_2$ and $M_3$.\boxtheorem
\end{example}}
\comlb{Need a better example instead of the above, where we get two different numbers and we keep the largest.}
\comlb{The next example should do. }
}

\begin{example}\label{ex:allNew} Consider instance $D = \{A(1;a), B(2;a), C(3;a), D(4;a), E(5;a)\}$, already with tids, and the set of DCs
$$\Sigma= \{\neg \exists x(B(x)\wedge E(x)), \ \neg \exists x(B(x) \wedge C(x) \wedge D(x)), \ \neg \exists x(A(x) \wedge C(x))\}.$$ The conflict hyper-graph whose hyper-edges connect tuples that simultaneously violate a DC can found in Figure \ref{fig:chg} \cite{lopatenko}. The S-repairs, which are maximal independent sets in the hyper-graph, are: \ $D_1 = \{B(a), C(a)\}$, \ $D_2 = \{C(a), D(a),$ $ E(a)\}$, \ $D_3 = \{A(a),B(a), D(a)\}$ and
$D_4 =\{E(a), D(a), A(a)\}$.

\begin{figure}[h]
\includegraphics[width=4cm]{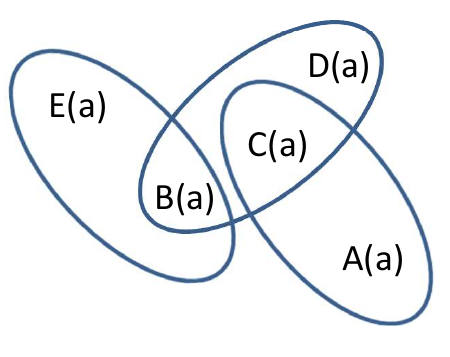}
\caption{Conflict hyper-graph}\label{fig:chg}
\end{figure}

The repair-program contains the repair-rules:
\begin{eqnarray*}
B'(t_1;x,\sfd) \vee E'(t_2;x,\sfd)&\leftarrow&B(t_1;x),E(t_2;x).\\
B'(t_1;x,\sfd) \vee C'(t_2;x,\sfd)\vee D(t_3;x)&\leftarrow&B(t_1;x),C(t_2;x),D(t_3;x).\\
A'(t_1;x,\sfd)\vee C'(t_2;x,\sfd)&\leftarrow&A(t_1;x),C(t_2;x).
\end{eqnarray*}
Its four stable models $M_1, M_2, M_3, M_4$ are in correspondence with the  S-repairs $D_1, D_2, D_3, D_4$.

Repairs $D_1$ and $D_2$ already sanction $A(a)$ as an actual cause for the query $\mc{Q}$ that is the disjunction of the negated DCs, i.e. a  UBCQs, namely

\centerline{$\mc{Q}\!: \ \exists x(B(x)\wedge E(x)) \vee \exists x(B(x) \wedge C(x) \wedge D(x)) \vee \exists x(A(x) \wedge C(x))$.}

 Furthermore, from $M_1$ and $M_2$, now as extended models for the causality-program obtained from the repair-program, we get answers $n = 2$ and $n=1$ to the query $:\!\!- \ \mbox{\nit{pre-rho}}(1,n)$ under the brave semantics, because in $M_1$ two tuples are deleted in addition to $A(1;a)$, but only one in $M_2$.  So, $n=1$ minimizes the query about $\mbox{\nit{pre-rho}}(1,n)$, with "$1$" here standing for the tid of $A(a)$. From the answer, the (global) responsibility $\rho(A(1;a)) = \frac{1}{2}$ can be obtained.

 We can see that, in order to specify causes for UBCQs, it is good enough to start from the single repair-program for the several DCs that correspond to the disjuncts in the query.
\boxtheorem
\end{example}

When dealing with a set of DCs, each repair rule of the form  (\ref{eq:repRule}) solves the corresponding local inconsistency, even if there is  interaction between the DCs, i.e. atoms in common, and other inconsistencies w.r.t. other DCs are solved at the same time. However, the minimal-model property of stable models makes sure that, in the end, a minimal set of atoms is deleted to solve all the inconsistencies \cite{monica}.

\red{The programs for the examples in this section are written and run with the DLV system in Appendix \ref{sec:dlv}.}

\section{ \ Specifying Attribute-Based Repairs and Causes}\label{sec:abc}

In this section we show how to specify repair-programs for null-based repairs, and how to extend and use them for causality specification and computation. The approach is general, but we use representative examples to convey it.

\begin{example} \label{ex:minimal} Consider the instance $D=\{\nit{Path}(1,2),\nit{Near}(2,1)\}$ for schema $\mc{R} =
\{\nit{Path}(\cdot,\cdot),$ $\nit{Near}(\cdot,\cdot)\}$. With tuple identifiers $8$ and $9$ it takes the form $D=\{\nit{Path}(8;1,2),$ $\nit{Near}(9;2,1)\}$.
Consider also the DC:\footnote{It would be easy to consider tids
in queries and view definitions, but they do not contribute to the final result and will only complicate the notation. So, we
skip tuple ids whenever possible.} \begin{equation}\kappa\!: \ \ \neg \exists x \exists y \exists z(\nit{Path}(x,y) \wedge \nit{Near}(y,z)), \label{eq:lastDC}\end{equation} which is violated by $D$.

Now,
consider the following alternative, updated instances $D_i$, each them obtained by replacing attribute values by \nn:

\begin{center}
 \begin{tabular}{|c|l|}
  \hline
  $D_1$ & $\{\nit{Path}(8;1,\nn),\nit{Near}(9;2,1)\}$ \\
  $D_2$ & $\{\nit{Path}(8;1,2),\nit{Near}(9;\nn,1)\}$ \\
  $D_3$ & $\{\nit{Path}(8;1,\nn),\nit{Near}(9;\nn,1)\}$ \\
  \cline{1-2}
 \end{tabular}
\end{center}

The sets of changes can be identified with the set of changed positions, as in Section \ref{sec:arepairs}, e.g. $\Delta^\nn(D,D_1)$ $ =
\{\nit{Path}[8;2]\}$ and $\Delta^\nn(D,D_2) = \{\nit{Near}[9;2]\}$ (remember that the tuple id goes always in position $0$). These $D_i$ are all consistent, but
$D_1$ and $D_2$ are the only null-based repairs of $D$; in particular they are $\leq_D^\nn$-minimal: The sets of changes $\Delta^\nn(D,D_1)$ and
$\Delta^\nn(D,D_2)$ are
  incomparable under set inclusion.
 $D_3$ is not $\leq_D^\nn$-minimal, because $\Delta^\nn(D,D_3) = \{\nit{Path}[8;2],\nit{Near}[9;2]\} \supsetneqq \Delta^\nn(D,D_2)$. \boxtheorem
\end{example}

 As in Section \ref{sec:specTuples}, null-based repairs can be specified as the stable models of a
repair-program, which we show next by means of Example \ref{ex:logicP}.
\ Actually, repair-programs for null-based repairs are inspired by ASP-programs that have been used to specify virtually and minimally updated versions of a database $D$ that is protected from revealing certain view contents \cite{tkde}. This is achieved by replacing direct query answering on $D$ by simultaneously querying (under the certain semantics) the virtual versions of $D$.

When we have more than one DC,  in contrast to the tuple-based repair semantics, where we can locally solve each inconsistency without considering  inconsistencies w.r.t. other DCs, a tuple that is subject to a local attribute-value update (into \nn), to solve one inconsistency, may need further updates to solve other inconsistencies. For example, if we add in Example \ref{ex:minimal} the DC

\centerline{$\kappa'\!: \ \neg \exists x \exists y (\nit{Path}(x,y) \wedge \nit{Near}(y,x))$,}

 \noindent the updates in repair $D_1$ have to be further continued, producing: $\nit{Path}(1;\nn,$ $\nn), \nit{Near}(2;\nn,$ $\nn)$. In other words, every locally updated tuple is considered to:  ``be in transition" or ``being updated" only (not necessarily in a definitive manner), until all inconsistencies are solved.

This observation motivates the introduction of the annotation constants that repair-programs will use now, for null-based repairs. The intended, informal semantics of annotation constants is
shown in Table \ref{table:annot}. The precise semantics is captured through the program that uses them.
As in Section \ref{sec:specTuples}, we use primed versions $R'$ of the original predicates $R$, and they will have tids and annotations as arguments as well.
\ignore{ More precisely, for each database predicate $R \in
\mc{R}$, we introduce a copy of it with  an extra, final attribute (or argument) that contains
an annotation constant. So, a tuple of the form $R(t;\bar{c})$ would become an annotated atom of the
form $R'(t;\bar{c},\mathbf{a})$.  }

\begin{table}[h]
\caption{General tuple annotations}
\label{table:annot}
\begin{tabular}{|l|l|l|}
 \hline
  Annotation & Atom & Tuple $R(\bar a)$ is ... \\
  \hline
 \hspace*{4mm}  $\au$ &  $R'(t;\bar{a}, \au)$ & ~the result of an update  \\
\hspace*{4mm}  $\bu$ &  $R'(t;\bar{a}, \bu)$ &  ~the final update of a tuple\\
\hspace*{4mm}  $\bft$ &  $R'(t;\bar{a}, \bft)$ & ~an initial or updated tuple  \\
 \hspace*{4mm} $\s$ & $R'(t;\bar{a}, \s)$ & ~definitive, to stay in the repair\\
  \hline
\end{tabular}
\end{table}

The  annotation constants are
used to keep track of virtual updates, i.e. of old and new tuples: \ An original tuple $R(t;\bar{c})$ may be successively updated, each time replacing an attribute value by \nn, creating tuples of
the form $R(t;\bar{c}',\au)$. Eventually the tuple will suffer no more updates, at which point it will become of the form $R'(t;\bar{c}'',\bu)$. In the transition, to check the satisfaction of the DCs,  it will be combined with other tuples, which can be updated versions of other tuples or tuples in the database that have never been updated. Both kinds of  tuples are uniformly annotated
with $R'(t',\bar{d},\bft)$, for being in transition. In this way, several, possibly interacting DCs can be handled. The tuples that eventually form a repaired version of the original database are those of the form
 $R'(t;\bar{e},\s)$, and are the final versions of the updated original tuples or the original tuples that were never updated.

In $R'(t;\bar{a}, \bu)$, annotation $\bu$ means that the atom with tid $t$ has reached its final update (during the program evaluation). In particular, $R(t;\bar{a})$ has already been updated, and $\au$
should appear in the new, updated atom, say $R'(t;\bar{a}',\au)$, and this tuple cannot be updated any further (because relevant updateable attribute values have already been replaced by \nn \ if necessary).
For example, consider a tuple $R(t;a,b)\in D$. A new tuple $R(t;a,\nn)$ is obtained by updating $b$ into $\nn$.
Therefore, $R'(t;a,\nn,\au)$ denotes the updated tuple. If this tuple is not updated any further, it will also eventually appear as $R'(t;a,\nn,\bu)$, indicating it is a final update.\footnote{Under null-based repairs no tuples are deleted or inserted, so the original tids stay all in the repairs and none is created.} \ (Cf. rules 3. in Example \ref{ex:logicP}.)

The repair-program uses these annotations to go through different steps, until its stable models are computed. Finally,
the atoms needed to build a repair are read off by restricting a model of the program to atoms with the annotation $\s$.
 The following example illustrates
the main ideas and issues.

\begin{example} (ex. \ref{ex:minimal} cont.) \label{ex:logicP} \ Consider   $D=$ $\{\nit{Path}(1,2),$ $\nit{Near}(2,1)\}$ and the DC: \linebreak \ $\kappa\!: \ \neg \exists x \exists y \exists z( \nit{Path}(x,y) \wedge \nit{Near}(y,z))$. \ The repair-program $\Pi(D, \{\kappa\})$ uses several auxiliary predicates to make the rules {\em safe}, i.e. with all their variables appearing in positive atoms in their bodies. It is as follows:

\vspace{2mm}
\hspace*{1mm} 1.~ \ \ \ \  $\nit{Path}(1;1,2)$. \ \ $\nit{Near}(2;2, 1).$ \ \ \ \  (initial database) \vspace{-2mm}
\begin{eqnarray*}
\mbox{\phantom{2mm}}2.\mbox{\phantom{ooo}}\nit{Path}'(t_1;x,\nn,\au) \vee \nit{Near}'(t_2;\nn,z,\au) &\leftarrow& \nit{Path}'(t_1;x,y,\bft),\\&&\hspace*{-4mm}\nit{Near}'(t_2;y,z,\bft), y\neq \nn.
\end{eqnarray*}

\vspace{-10mm}\begin{eqnarray*}
3.\mbox{\phantom{oooo}}\nit{Path}'(t;x,y,\bu)\! &\leftarrow& \nit{Path}'(t;x,y,\au), \nit{not} \ \nit{aux}_{\nit{Path}.1}(t;x,y),\\&& ~~ \nit{not} \ \nit{aux}_{\nit{Path}.2}(t;x,y). \\
\nit{aux}_{\nit{Path}.1}(t;x,y) &\leftarrow& \nit{Path}'(t;\nn,y,\au), \nit{Path}(t;x,z), x\neq \nn.\\
\nit{aux}_{\nit{Path}.2}(t;x,y) &\leftarrow& \nit{Path}'(t;x,\nn,\au), \nit{Path}(t;z,y), y\neq \nn.\\&&\hspace*{4.5cm}\mbox{\small (idem for $\nit{Near}$)}\\
4.\mbox{\phantom{ooooo}}\nit{Path}'(t;x,y,\bft) &\leftarrow& \!\nit{Path}(t;x,y). \\ \nit{Path}'(t;x,y,\bft) &\leftarrow& \nit{Path}'(t;x,y,\au).\hspace{2cm} \mbox{\small (idem for $\nit{Near}$)}\\
5.\mbox{\phantom{ooooo}}\nit{Path}'(t;x,y,\mathbf{s}) &\leftarrow& \!\nit{Path}'(t;x,y,\bu). \hspace{2cm} \mbox{\small (idem for $\nit{Near}$)}\\
        \nit{Path}'(t;x,y,\mathbf{s}) &\leftarrow& \!\nit{Path}(t;x,y),~\nit{not} \ \nit{aux}_\nit{Path}(t).\\
        \nit{aux}_\nit{Path}(t) &\leftarrow& \nit{Path}'(t;u,v,\au).
        \end{eqnarray*}

In this repair-program tids in rules are handled as variables; and constant \nn \ in the program is treated as any other constant. The latter is the reason for the condition $y \neq \nn$ in the body of 2., to avoid considering the join through \nn \ a violation of the DC.\footnote{If instead of (\ref{eq:lastDC}) we had $\kappa\!: \ \neg \exists x \exists y \exists z( \nit{Path}(x,y) \wedge \nit{Near}(y,z) \wedge y<3)$, the new rule body  could be $\nit{Path}'(t_1;x,y,\bft),~ \nit{Near}'(t_2;y,z,\bft),y < 3$, because $\nn < 3$ would be evaluated as false.}
 A quick look at the program shows that the original tids are never destroyed and no new tids are created, which simplifies keeping track of tuples under repair updates. It also worth mentioning that for this particular example, with a single DC, a much simpler program could be used, but we keep the general form that can be applied to multiple, possibly interacting DCs.

Facts  in 1. belong to the initial instance $D$, and become annotated right away with $\bft$ by rules 4. The most
important rules of the program are those in  2. They enforce one step of the update-based repair-semantics  in the presence of
$\nit{null}$ and
using $\nit{null}$ (right,  already having nulls in the initial database is not a problem). Rules in 2. capture in the body the violation of DC; and in the head, the intended way of
restoring consistency, namely making  one of the attributes participating in a join take value \nn.

Rules in 3. collect the final updated versions  of the tuples in the database, as those whose values are never replaced by a \nn \ in another updated version. \ignore{The auxiliary predicates are used to have {\em safe rules}, i.e. to avoid having negation in front of atoms with variables that do not appear in a positive atom in the same body. Notice that the conditions
$x \neq \nn$ and $y\neq \nn$ in bodies of the definitions of the auxiliary predicates may be replaced by $\nit{Path}(t;x,z), x\neq \nn$ and $\nit{Path}(t;z,y), y\neq \nn$, resp., if we want variables $x$ and $y$ to range over a bounded domain.} \
Rules in 4. annotate the original atoms and also new versions of updated atoms. They all can be subject of additional updates and have to be checked for DC satisfaction, with rule 2.. Rules in 5. collect the tuples
that stay in the final state of the updated database, namely the original and never updated tuples plus the final, updated versions of tuples. \boxtheorem
\end{example}

\begin{proposition} \em There is a one-to-one correspondence between the \nn-based repairs of $D$ w.r.t. a set of DCs $\Sigma$ and the stable models of the repair-program $\Pi(D, \Sigma)$. More specifically, a repair $D'$ can be obtained
by collecting the $\mathbf{s}$-annotated atoms in a stable
model $M$, i.e. \ $D' = \{P(\bar{c})~|~ P'(t;\bar{c},\mathbf{s}) \in M\}$; and every repair can be obtained in this way.\footnote{The proof of this claim is relatively straightforward, but rather long. It follows the same pattern  as
 the proof that tuple-based repairs w.r.t. integrity constraints
can be specified
by means of disjunctive logic programs with stable model semantics  \cite[prop. 6.1]{tplp17}.} \boxtheorem
\end{proposition}

\begin{example}\label{ex:logicP2} (ex. \ref{ex:logicP} cont.) The program has two stable models: \ (the facts in 1. and the \nit{aux}-atoms
are omitted)
\begin{eqnarray*}
M_1 &=& \{\nit{Path}'(1;1,2,\bft),~ \nit{Near}'(2;2,1,\bft),\underline{\nit{Near}'(2;2,1,\s)},\nit{Path}'(1;1,\nn,\au),\\ &&
~~~~~~~~~~\nit{Path}'(1;1,\nn,\bft),\nit{Path}'(1;1,\nn,\bu), \underline{\nit{Path}'(1;1,\nn,\s)}\}.\\
M_2 &=&\{\nit{Path}'(1;1,2,\bft),~ \nit{Near}'(2;2,1,\bft), \underline{\nit{Path}'(1;1,2,\s)}, \nit{Near}'(2;\nn,1,\au),\\ &&
~~~~~~~~~~\nit{Near}'(2;\nn,1,\bft),\nit{Near}'(2;\nn,1,\bu),\underline{\nit{Near}'(2;\nn,1,\s)}\}.
\end{eqnarray*}
The repairs  are built by selecting the underlined atoms: $D_1$ $=$ $\{\nit{Path}(1,\nn),$ $\nit{Near}(2,1)\}$ and $D_2 =
\{\nit{Path}(1,2), \nit{Near}(\nn,1)\}$. They coincide with those in Example \ref{ex:minimal}.
\boxtheorem
\end{example}

\red{Appendix \ref{sec:B} shows, in Example \ref{ex:veryLast}, a program written in DLV for null-based causes.}

 Finally, and similarly to the use of repair-programs for cause computation in Section \ref{sec:specTuples}, we can extend the new repair-programs into causality-programs, to compute attribute-null-based causes (we do not consider here tuple-null-based causes, nor the computation of responsibilities, all of which can be done along the lines of Section \ref {sec:specTuples}). All we need to do is add to the repair-program the definition of a cause predicate, through rules of the form: $$\nit{Cause}(t;i;v) \leftarrow R'(t;\bar{x},\nn,\bar{z},\s), R(t;\bar{x}',v,\bar{z}'), v \neq \nn,$$ (with $v$ and \nn \ in the body in the same position $i$), saying that  value $v$ in the $i$-th position in original tuple with tid $t$ is an attribute-null-based cause. The condition $v \neq \nn$ can be skipped in general since it is useful only in case the original instance already has null values. The rule collects the original values (with their tids and positions) that have been changed into \nn.

 In Example \ref{ex:logicP} we would add to the repair-program the rules \  (with similar rules for predicate $\nit{Near}$) \begin{eqnarray*}\nit{Cause}(t;1;x) &\leftarrow& \nit{Path}'(t;\nn,y,\s), \nit{Path}(t;x,y').\\ \nit{Cause}(t;2;y) &\leftarrow& \nit{Path}'(t;x,\nn,\s), \nit{Path}(t;x',y).
 \end{eqnarray*}
The rules for contingency set and responsibility computation are as for tuple-based causes in Section \ref{sec:specTuples}.


\section{ \ Causes under Integrity Constraints}\label{sec:ics}

   For query-answer causality in databases, taking into account ICs that are expected to be satisfied by the database becomes natural. In fact, the problem of characterizing and computing causes for query answers in the presence of ICs was investigated in \cite{flairsExt}. We now briefly recall the main notions involved.

   Assume we have a set $\Psi$ of ICs, and a database $D$, and $D \models \Psi$. This set of constraints is assumed to be {\em hard}, in that their violation is never acceptable.
 \ The definition of  $\tau \in D$ \ as an actual cause for $\mc{Q}(\bar{a})$ in $D$ and wrt. $\Psi$ is as in Section \ref{sec:tcause},  but now for the contingency set $\Gamma$ we require:
\ ${D \smallsetminus \Gamma \ \models \ \Psi}$, \ $D \smallsetminus \Gamma \ \models \ \mc{Q}(\bar{a})$, \
$D \smallsetminus (\Gamma \cup \{\tau\}) \ \models \ \Psi$, and  $D \smallsetminus (\Gamma \cup \{\tau\}) \ \not \models \ \mc{Q}(\bar{a})$. \ Responsibility, denoted \ $\rho_{_{\!\mc{Q}(\bar{a})\!}}^{D,\Psi}(\tau)$, \ is defined  as before.

\begin{example} \label{ex:wICS} Consider the  instance $D$ and the inclusion dependency $$\psi: \  \ \forall x \forall y \ (\nit{Dep}(x, y) \rightarrow \exists u  \  \nit{Course}(u, y)),$$ which is a non-monotonic IC, and is satisfied by $D$.\footnote{It is non-monotonic in that its violation view, which captures the tuples that violate it, is defined by a non-monotonic query. Monotonic ICs, i.e. for which a growing database may only produce more violations (e.g. denial constraints and FDs),  are not much of an issue in this causality setting with conjunctive queries, because they stay satisfied under counterfactual deletions associated to causes. So here, the most relevant of the usual ICs are non-monotonic.}

\begin{center}
{\scriptsize \begin{tabular}{c|c|c|} \hline
\nit{ Dep} & \nit{DName} &\nit{TStaff}  \\\hline
$\tau_1$& {\sf computing} & {\sf john}  \\
$\tau_2$& {\sf philosophy} &  {\sf patrick}   \\
$\tau_3$&{\sf math}  &  {\sf kevin}   \\
 \hhline{~--} \end{tabular}~~~~~~~~~~
 \begin{tabular}{c|c|c|} \hline
\nit{Course}  & \nit{CName} & \nit{TStaff} \\\hline
$\tau_4$&{\sf com08} & {\sf john}  \\
$\tau_5$&{\sf math01} & {\sf kevin} \\
$\tau_6$&{\sf hist02}&  {\sf patrick} \\
$\tau_7$&{\sf math08}&  {\sf eli}\\
$\tau_8$&{\sf com01}&  {\sf john}\\
 \hhline{~--}
\end{tabular} }
 \end{center}

Consider the query \ $\mc{Q}_1({x})\!: \ \exists y \exists z (\nit{Dep}(y,{x}) \wedge
\nit{Course}(z,x))$, for which   $\langle{\sf john}\rangle \in \mc{Q}_1(D)$. \ Without considering $\psi$:
(a) \ $\tau_1$ is a counterfactual cause; \ (b) \
$\tau_4$ is actual cause with minimal contingency set $\Gamma_1=\{\tau_8\}$; \ (c) \
$\tau_8$ \ is actual cause with minimal contingency set \ $\Gamma_2=\{\tau_4\}$.
\ However, under $\psi$,  $\tau_4$ and $\tau_4$ \ are not  actual causes  anymore; but $\tau_1$ is still is counterfactual cause.

Now consider the query \  $\mc{Q}_2({x})\!: \ \exists z \nit{Course}(z,{x})$, for which   $\langle{\sf john}\rangle \in \mc{Q}_2(D)$. \ Without $\psi$,
  $\tau_4$ and $\tau_8$ are the only actual causes, with minimal contingency sets  $\Gamma_1 = \{\tau_8\}$ and $\Gamma_2 = \{\tau_4\}$, resp.

If we consider $\psi$,  $\tau_4$ and $\tau_8$ are still  actual causes, but we  lose $\Gamma_1$ and $\Gamma_2$ as contingency sets. Actually, the
smallest contingency set for $\tau_4$ is $\Gamma_3 = \{\tau_8, \tau_1\}$, and for  $\tau_8$, it is $\Gamma_4 = \{\tau_4, \tau_1\}$. Accordingly, the
responsibilities of \ $\tau_4, \ \tau_8$ \ decrease: \  {$\rho_{_{\mc{Q}_2({\sf john})}}^D(\tau_4) = \frac{1}{2}$, but  $\rho_{_{\mc{Q}_2({\sf john})}}^{D,\psi}(\tau_4) =\frac{1}{3}$}. Notice that
$\tau_1$ is still not an actual cause, but it affects the responsibility of actual causes. \boxtheorem
\end{example}

From \cite{flairsExt}, we know that causes are preserved under logical equivalence of queries under ICs, and that deciding causality for conjunctive queries under inclusion dependencies can be NP-complete (the same problem  is tractable without ICs).

\begin{example} \label{ex:wICS2} (ex. \ref{ex:wICS} cont.) \ Database $D$ violates the DC \ $\kappa_2: \ \neg \exists z \nit{Course}(z,$ ${\sf John})$ \ associated to query $\mc{Q}_2$ and its  answer {\sf John}. Without considering $\psi$, its only minimal repair is $D' = D \smallsetminus \{\tau_4,\tau_8\}$. However, if we accept minimal repairs that also satisfy $\psi$ (when $D$ already did so), then the only minimal repair is $D'' = D \smallsetminus \{\tau_1, \tau_4,\tau_8\}$. \boxtheorem
\end{example}
This example shows that, in the presence of a set of hard ICs $\Psi$, the repairs wrt. to another set of ICs $\Sigma$ that also satisfy $\Psi$ may not be among the repairs wrt. $\Sigma$ without consideration for $\Psi$. So, it is not only a matter of discarding some of the unwanted repairs wrt. $\Sigma$ alone.

 The example also shows that, in the presence of a hard set of ICs $\Psi$, the characterization of causes in terms of repairs  and
 Proposition \ref{prop:tocs} have to be revised. \ Doing this should be relatively straightforward for repairs of $D$ w.r.t.  the DCs $\Sigma$ that have origin in the UBCQs, and are maximally contained in $D$ under set-inclusion, and also satisfy the hard constraints $\Psi$. Instead of doing this, we show how a repair-program could be used to reobtain the results obtained in Example \ref{ex:wICS}.

\begin{example} (exs. \ref{ex:wICS} and \ref{ex:wICS2} cont.)  Without considering the IC $\psi$, the repair-program for $D$ wrt. the DC $\kappa_2$ is:
\begin{enumerate}
\item The extensional database as a set of facts corresponding to the table. For example, \ $\nit{Dept}(1;{\sf computing},{\sf john})$, etc.

\item Repair rule for $\kappa_2$: \ \ \ $\nit{Course}'(t;z,{\sf john},\sfd) \leftarrow \nit{Course}(t;z,{\sf john}).$

\item Persistence rule: \ \ \ $\nit{Course}'(t;x,y,\sfs) \leftarrow \nit{Course}(t;x,y), \ \nit{not} \ \nit{Course}'(t;x,y,\sfd).$
\end{enumerate}
To this program we have to add rules that take care of repairing w.r.t. $\psi$ in case it is violated via deletions from $\nit{Course}$:
  \begin{enumfrom}{4}
\item    $\nit{Dept}'(t',x,y,\sfd)  \leftarrow \nit{Dept}(t',x,y), \nit{not} \ \nit{aux}(y)$

\item $\nit{aux}(y) \leftarrow \nit{Course}'(t;x,y,\sfs)$.

\item $\nit{Dept}'(t;x,y,\sfs) \leftarrow \nit{Dept}(t;x,y), \ \nit{not} \ \nit{Dept}'(t;x,y,\sfd).$
\end{enumfrom}
Notice that violations of  the inclusion dependency that may arise from deletions from $\nit{Course}$ are being repaired through deletions from $\nit{Dept}$. \ The only stable model of this program corresponds to the repair in Example \ref{ex:wICS2}. \boxtheorem
\end{example}

Notice that the definition of actual cause under ICs opens the ground for a definition of a notion of {\em underlying} ({\em hidden}, {\em latent}) cause. In Example \ref{ex:wICS}, $\tau_1$ could be such a cause. It is not strictly an actual cause, but it has to appear in every minimal contingency set. Similarly, Example \ref{ex:wICS2} shows that $\tau_1$ has to appear in the difference between the original instance and every minimal repair. We leave this extension and its analysis for future work.

\section{ \ Discussion}\label{sec:disc}
\paragraph{Complexity.} \ Computing causes for CQs can be done in polynomial time in data \cite{Meliou2010a}, which also holds for  UBCQs \cite{tocs}. In \cite{flairsExt} it was established that cause computation for Datalog queries falls in the second level of the polynomial hierarchy (PH). As has been established in \cite{Meliou2010a,tocs}, the computational problems associated to contingency sets and responsibility are on the second level of PH, in data complexity.

 On the other side, our repairs programs, and so our causality-programs, i.e. repair-programs with causality extensions, can be transformed into non-disjunctive, unstratified programs \cite{barcelo,monica}, whose reasoning tasks are also on the second level of PH (in data complexity) \cite{dantsin}. It is worth mentioning that the ASP approach to causality via repair-programs could be extended to deal with queries that are more complex than CQs or UCQs, e.g. Datalog queries and  queries that are conjunctions of literals (that were investigated in \cite{tapp16}). This corresponds to ongoing work.

\ignore{++
\vspace{-2mm}
\paragraph{Causality programs and ICs} \ The original causality setting in \cite{Meliou2010a} does not consider ICs. An extension of causality under ICs was proposed in \cite{flairsExt}. Under it,  the ICs have to be satisfied by the databases involved, i.e. the initial one and those obtained by cause and contingency-set deletions. When the query at hand is monotonic,\footnote{I.e. the set of answers may only grow when the instance grows.} monotonic ICs, i.e. for which growing with the database may only produce more violations (e.g. denial constraints and FDs),  are not much of an issue since they stay satisfied under deletions associated to causes. So, the most relevant ICs are non-monotonic, such as inclusion dependencies, e.g. $\forall xy(R(x,y) \rightarrow S(x))$. These ICs can be represented in a causality-program by means of (strong) program constraints. In the running example, we would have, for tuple-based causes, the
constraint: \ $\leftarrow R'(t,x,y,\sfs), \nit{not} \ S'(t',x,\sfs)$.\footnote{Or better, to make it {\em safe}, by a rule and a constraint: \ $\nit{aux}(x) \leftarrow S'(t',x,\sfs)$ and \linebreak $\leftarrow R'(t,x,y,\sfs), \nit{not}
\ \nit{aux}(x)$.}
++}

\vspace{-2mm}\paragraph{Negative CQs and inclusion dependencies} \ In this work we investigated CQs, and what we did can be extended to UCQs. However, it is possible to consider queries that are conjunctions of literals, i.e. atoms or negations thereof, e.g. $\mc{Q}\!: \ \exists x\exists y(P(x,y) \wedge \neg S(x))$.\footnote{They should be {\em safe} in the sense that a variable in a negative literals has to appear in some positive literal too.} Causes for these queries were investigated in \cite{tapp16}. If causes are defined in terms of counterfactual deletions (as opposed to insertions that can also be considered for these queries), then the repair counterpart can be constructed by transforming the query into the unsatisfied  {\em inclusion dependency} (ID): $\forall x \forall y(P(x,y) \rightarrow S(x))$. Repairs w.r.t. this kind of IDs that allow only tuple deletions were considered in \cite{chomicki}, and repairs programs for them in \cite{monica}. Causes for CQs in the presence of IDs were considered in \cite{flairsExt}. Actually, Example \ref{ex:wICS2} shows this approach in that only deletions are used to restore consistency wrt. the inclusion dependency.

\vspace{-2mm}
\paragraph{Endogenous  and prioritized causes and repairs.} \ As indicated in Section \ref{sec:abstract}, different kinds of causes can be introduced by considering different repair-semantics. Apart from those investigated in this work, we could consider {\em endogenous repairs}, which are obtained by removing only  endogenous tuples \cite{tocs}. In this way we could give an account of causes as in Section \ref{sec:tcause}, but considering the partition of the database between endogenous and exogenous tuples.

Again, considering the abstract setting of Section  \ref{sec:abstract}, with  the generic class of repairs \
  $\nit{Rep}^{{\sf S}^\preceq\!}(D,\Sigma)$,  it is possible to consider different kinds of {\em prioritized repairs} \cite{stawo}, and through them introduce {\em prioritized actual causes}.  Repair-programs for the kinds of priority relations $\preceq$ investigated in \cite{stawo} could be constructed with the ASPs introduced and investigated in \cite{gebser} for capturing different optimality criteria. The repair-programs could be used, as done in this work, to specify and compute the corresponding prioritized actual causes and responsibilities.

\vspace{-2mm}
\paragraph{Qualitative responsibilities.} The abstract definition of an actual cause on the basis of an also abstract repairs semantics (cf. Definition \ref{def:absCause}) opens the ground for defining a qualitative, preference-based notion of responsibility.
Priorities and preferences on tuples could be considered when bringing tuples into a an actual cause's  contingency set. The ``better" the tuples in a contingency set, the better the actual cause for the query result. This idea deserves investigation.

\vspace{-2mm}
\paragraph{Optimization of causality programs.} \ Different queries  about causality could be posed to our causality-programs or directly to the underlying repair-programs. Query answering could benefit from query-dependent, magic-set-based optimizations of causality and repair-programs as reported in \cite{monica}. Implementation and experimentation in general are left for future work.

\vspace{-2mm}
\paragraph{Connections to Belief Revision/Update.} \ As discussed in \cite{pods99} (cf. also \cite{Bertossi2011}), there are some connections between database repairs and belief updates as found in knowledge representation, most prominently with \cite{winslett}.  \ In \cite{tplp}, some connections were established between repair-programs and {\em revision programs} \cite{marek}. The applicability of the latter in a causality scenario like ours becomes a matter of future investigation.

\begin{acknowledgements} Part of this work was done by the author as a Senior Computer Scientist at {\em RelationalAI Inc.}, and also
while the author was spending a sabbatical at the ``Database and Artificial Intelligence" Group of the Technical University of Vienna with support from the ``Vienna Center for Logic and Algorithms" and the Wolfgang Pauli Society. The author is extremely grateful for their support and hospitality, and especially to Prof. Georg Gottlob for making the stay possible. \ Many thanks to the anonymous reviewers \cite{foiks18} for their excellent feedback. \ The valuable help provided by Jordan Li with the examples in DLV is much appreciated. \ Part of this work was funded by ANID - Millennium Science Initiative Program - Code ICN17\_002.
\end{acknowledgements}



\appendix

\section{ \ \red{Examples of Tuple-Based Causes with DLV-Complex}}\label{sec:dlv}

In this section we show in detail how the examples and repairs-programs extended with causality elements of Section \ref{sec:specTuples} can be specified and executed  in the DLV-Complex system \cite{calimeri08,calimeri09}.

\begin{example} \label{ex:rep1} (ex. \ref{ex:cause}, \ref{ex:kappa}, \ref{ex:kappa2}-\ref{ex:wcs} cont.) \ The first fragment of the DLV  program below, in its non-disjunctive version, shows facts for  database $D$, and the three repair rules for the DC $\kappa(\mc{Q})$. In it, and in the rest of this section, predicates \verb+R+, \verb+S+ stand for $\R$ and $\St$, resp.; and \verb+R_a, S_a, ...+ stand for $\R', \St', ...$ used before, with the subscript \verb+_a+ for ``auxiliary". We recall that the first attribute of a predicate holds a variable or a constant for a tid; and the last attribute of \verb+R_a+, etc. holds an annotation constant, \verb+d+ or \verb+s+, for ``deleted" (from the database) or ``stays" in a repair, resp.

{\footnotesize
\begin{verbatim}
    R(1,a4,a3). R(2,a2,a1). R(3,a3,a3). S(4,a4). S(5,a2). S(6,a3).

    S_a(T,X,d)   :- S(T,X), R(T2,X,Y), S(T3,Y), not R_a(T2,X,Y,d),
                    not S_a(T3,Y,d).
    S_a(T,X,d)   :- S(T,X), R(T2,Y,X), S(T3,Y), not R_a(T2,Y,X,d),
                    not S_a(T3,Y,d).
    R_a(T,X,Y,d) :- R(T,X,Y), S(T2,X), S(T3,Y), not S_a(T2,X,d),
                    not S_a(T3,Y,d).
    S_a(T,X,s)   :- S(T,X), not S_a(T,X,d).
    R_a(T,X,Y,s) :- R(T,X,Y), not R_a(T,X,Y,d).
\end{verbatim} }

This is the non-disjunctive version of the repair-program given in disjunctive form in Example \ref{ex:kappa2}, which in DLV takes the following form: (we will keep using the non-disjunctive versions of these programs)
\ignore{Equivalently, the constraint can be enforced using a disjunctive rule. The following code fragment defines the exact same EDB and constraint as the previous code by using a single disjunctive rule to represent the DC. While the two versions have identical functions, we will present the non-disjunctive versions for further code examples.}

{\footnotesize \begin{verbatim}
    R(1,a4,a3). R(2,a2,a1). R(3,a3,a3). S(4,a4). S(5,a2). S(6,a3).

    S_a(T1,X,d) v R_a(T2,X,Y,d) v S_a(T3,Y,d) :- S(T1,X),R(T2,X,Y), S(T3,Y).
    S_a(T,X,s)   :- S(T,X), not S_a(T,X,d).
    R_a(T,X,Y,s) :- R(T,X,Y), not R_a(T,X,Y,d).
\end{verbatim} }
DLV returns the stable models of the program, as follows:
{\footnotesize
\begin{verbatim}
    {S_a(4,a4,d), R_a(3,a3,a3,d), R_a(1,a4,a3,s), R_a(2,a2,a1,s),
     S_a(5,a2,s), S_a(6,a3,s)}

    {R_a(1,a4,a3,d), R_a(3,a3,a3,d), R_a(2,a2,a1,s), S_a(4,a4,s),
     S_a(5,a2,s), S_a(6,a3,s)}

    {S_a(6,a3,d), R_a(1,a4,a3,s), R_a(2,a2,a1,s), R_a(3,a3,a3,s),
     S_a(4,a4,s), S_a(5,a2,s)}
\end{verbatim} }
These three stable models (that do not show here the original EDB) are associated to the S-repairs $D_1, D_2, D_3$ in Example \ref{ex:kappa}, resp. As expected from Example \ref{ex:cause}, only tuples with tids $1,3,4,6$ are at some point deleted. In particular,
the last model corresponds to the C-repair \

\centerline{ $D_1 = \{\R(s_4,s_3),  \R(s_2,s_1), \R(s_3,s_3),$ $ \St(s_4), \St(s_2)\}$.}

\vspace{1mm}Now, to compute causes and their accompanying deleted tuples we add  to the program the  rules defining $\nit{Cause}$, in (\ref{eq:cause1})-(\ref{eq:cause2}), and $\nit{CauCont}$, in Example \ref{ex:causes}:

{\footnotesize
\begin{verbatim}
         cause(T) :- S_a(T,X,d).
         cause(T) :- R_a(T,X,Y,d).
    cauCont(T,TC) :- S_a(T,X,d), S_a(TC,U,d), T != TC.
    cauCont(T,TC) :- R_a(T,X,Y,d), R_a(TC,U,V,d), T != TC.
    cauCont(T,TC) :- S_a(T,X,d), R_a(TC,U,V,d).
    cauCont(T,TC) :- R_a(T,X,Y,d), S_a(TC,U,d).
\end{verbatim}}

Next, contingency sets can be computed by means of DLV-Complex, on the basis of the rules defining predicates $\nit{cause}$ and $\nit{cauCont}$ above:

{\footnotesize \begin{verbatim}
              preCont(T,{TC}) :- cauCont(T,TC).
    preCont(T,#union(C,{TC})) :- cauCont(T,TC), preCont(T,C), not #member(TC,C).
                  cont(T,C)   :- preCont(T,C), not HoleIn(T,C).
                  HoleIn(T,C) :- preCont(T,C), cauCont(T,TC), not #member(TC,C).
                  tmpCont(T)  :- cont(T,C), not #card(C,0).
                  cont(T,{})  :- cause(T), not tmpCont(T).
\end{verbatim}}

The last two rules play the role of rules (\ref{eq:cont3}) and (\ref{eq:cont4}), that associate the empty contingency set to counterfactual causes.\ignore{
\red{This code is the direct translation of the causality rules introduced in Section \ref{sec:specTuples} as well as $\nit{tmpCont}(T)$ and $\nit{cont}(T,\{\})$ which will give us empty contingency sets for counterfactual causes, for the sake of consistency with other causes.} }

The three stable models obtained above will now be extended with $\nit{cause}$- and $\nit{cont}$-atoms, among others (unless otherwise stated, $\nit{preCont}$-, $\nit{tmpCont}$-, and $\nit{HoleIn}$-atoms will be filtered out from the output); as follows:

{\footnotesize \begin{verbatim}
    {S_a(4,a4,d), R_a(3,a3,a3,d), R_a(1,a4,a3,s), R_a(2,a2,a1,s),
     S_a(5,a2,s), S_a(6,a3,s), cause(4), cause(3), cauCont(4,3),
     cauCont(3,4), cont(3,{4}), cont(4,{3})}

    {R_a(1,a4,a3,d), R_a(3,a3,a3,d), R_a(2,a2,a1,s), S_a(4,a4,s),
     S_a(5,a2,s), S_a(6,a3,s), cause(1), cause(3), cauCont(1,3),
     cauCont(3,1), cont(1,{3}), cont(3,{1})}

    {S_a(6,a3,d), R_a(1,a4,a3,s), R_a(2,a2,a1,s), R_a(3,a3,a3,s),
     S_a(4,a4,s), S_a(5,a2,s), cause(6), cont(6,{})}
\end{verbatim}}

The first two models above show tuple 3 as an actual cause, with one contingency set per each of the models where it appears as a cause. The last line of the third model shows that cause (with tid) 6 is the only counterfactual  cause (its contingency set is empty).

The responsibility $\rho$ can  be computed via predicate
$\nit{preRho}(T,N)$, defined in (\ref{eq:prho}), that returns  $N = \frac{1}{\rho}$, that is the inverse of the responsibility, for each tuple with tid $T$ {\em and local to a model} that shows $T$ as a cause. We concentrate on the computation of $\nit{preRho}$ in order to compute with integer numbers, as supported by DLV-Complex,\ignore{ only handles integers, thus we will work only with the integer denominator of the responsibility. Consequently, with this rule we see the tuple with the lowest pre-rho value as being the most responsible cause.} which requires setting an upper integer bound by means of \verb+maxint+, in this case, at \ignore{ must be set either in code or by using the "-N" command line option and must be at} least as large as the largest tid:

{\footnotesize \begin{verbatim}
    #maxint = 100.
    preRho(T,N + 1) :- cause(T), #int(N), #count{TC: cauCont(T,TC)} = N.
\end{verbatim}}

\noindent where the local (pre)responsibility of a cause (with tid) $T$ within a repair is obtained by counting how many instances of $\nit{cauCont}(T,?)$ exist in the model, which is the size of the local contingency set for $T$ plus 1. We obtain the following (filtered) output:

{\footnotesize \begin{verbatim}
    {S_a(4,a4,d), R_a(3,a3,a3,d), cause(4), cause(3),
     preRho(3,2), preRho(4,2), cont(3,{4}), cont(4,{3})}

    {R_a(1,a4,a3,d), R_a(3,a3,a3,d), cause(1), cause(3),
     preRho(1,2), preRho(3,2), cont(1,{3}), cont(3,{1})}

    {S_a(6,a3,d), cause(6), preRho(6,1), cont(6,{})}
\end{verbatim}}

The first model shows causes 3 and 4 with a pre-rho value of $2$. The second one, causes 3 and 1 with a pre-rho value of $2$. The last model shows cause 6 with a pre-rho value of $1$. This is also a maximum-responsibility cause, actually associated to a C-repair. Inspecting the three models, we can see that the overall pre-responsibility of cause 3 (the minimum of its pre-rho values) is $2$, similarly for cause 1. For cause 6 the overall pre-responsibility value is $1$.

Now, if we want only maximum-responsibility causes, we add weak program constraints to the program above,  to minimize the number of deletions:

{\footnotesize \begin{verbatim}
    :~ S_a(T,X,d).
    :~ R_a(T,X,Y,d).
\end{verbatim} }
\noindent DLV shows only repairs with the least number of deletions, in this case:
{\footnotesize \begin{verbatim}
    Best model: {S_a(6,a3,d), R_a(1,a4,a3,s), R_a(2,a2,a1,s), R_a(3,a3,a3,s),
                 S_a(4,a4,s), S_a(5,a2,s), cause(6), preRho(6,1), cont(6,{})}
    Cost ([Weight:Level]): <[1:1]>
\end{verbatim} }

As expected, only repair $D_1$ is obtained, where only $\St(6,s_3)$ is a cause, and with  responsibility $1$, making it a maximum-responsibility cause.
\boxtheorem
\end{example}


\begin{example} (ex. \ref{ex:allNew} cont.) \ We proceed as  in Example \ref{ex:rep1}, with the repair-program being: \ (again, in non-disjunctive form, but DLV accepts disjunction)
\ignore{We first add the facts for the database $D = \{A(1,a), \\ B(2,a), C(3,a), D(4,a), E(5,a)\}$ and repair rules based on the constraints $DC: \\ \neg \exists x(B(x)\wedge E(x)) \wedge \neg \exists x(B(x) \wedge C(x) \wedge E(x)) \wedge \neg \exists x(A(x) \wedge C(x))$. We write the program in DLV without disjunction as follows. As with the previous example, each predicate contains an extra attribute for unique tuple ids.}

{\footnotesize \begin{verbatim}
    A(1,a). B(2,a). C(3,a). D(4,a). E(5,a).

    B_a(T,X,d) :- B(T,X), E(T2,X), not E_a(T2,X,d).
    E_a(T,X,d) :- E(T,X), B(T2,X), not B_a(T2,X,d).

    B_a(T,X,d) :- B(T,X), C(T2,X), D(T3,X), not C_a(T2,X,d), not D_a(T3,X,d).
    C_a(T,X,d) :- C(T,X), B(T2,X), D(T3,X), not B_a(T2,X,d), not D_a(T3,X,d).
    D_a(T,X,d) :- D(T,X), B(T2,X), C(T3,X), not B_a(T2,X,d), not C_a(T3,X,d).

    A_a(T,X,d) :- A(T,X), C(T2,X), not C_a(T2,X,d).
    C_a(T,X,d) :- C(T,X), A(T2,X), not A_a(T2,X,d).

    A_a(T,X,s) :- A(T,X), not A_a(T,X,d).
    B_a(T,X,s) :- B(T,X), not B_a(T,X,d).
    C_a(T,X,s) :- C(T,X), not C_a(T,X,d).
    D_a(T,X,s) :- D(T,X), not D_a(T,X,d).
    E_a(T,X,s) :- E(T,X), not E_a(T,X,d).
\end{verbatim} }

Now we define the contingency sets and the local (pre)responsibilities for every cause in each model:

{\footnotesize \begin{verbatim}
    cause(T) :- A_a(T,X,d).
    cause(T) :- B_a(T,X,d).
    cause(T) :- C_a(T,X,d).
    cause(T) :- D_a(T,X,d).
    cause(T) :- E_a(T,X,d).

    cauCont(T,TC) :- A_a(T,X,d), A_a(TC,Y,d), T != TC.
    cauCont(T,TC) :- A_a(T,X,d), B_a(TC,Y,d).
    cauCont(T,TC) :- A_a(T,X,d), C_a(TC,Y,d).
    cauCont(T,TC) :- A_a(T,X,d), D_a(TC,Y,d).
    cauCont(T,TC) :- A_a(T,X,d), E_a(TC,Y,d).

    cauCont(T,TC) :- B_a(T,X,d), A_a(TC,Y,d).
    cauCont(T,TC) :- B_a(T,X,d), B_a(TC,Y,d), T != TC.
    cauCont(T,TC) :- B_a(T,X,d), C_a(TC,Y,d).
    cauCont(T,TC) :- B_a(T,X,d), D_a(TC,Y,d).
    cauCont(T,TC) :- B_a(T,X,d), E_a(TC,Y,d).
    ...
\end{verbatim} }

We obtain the following output, showing four S-repairs, with the last three being C-repairs:

{\footnotesize \begin{verbatim}
    {D_a(4,a,d), C_a(3,a,s), A_a(1,a,d), E_a(5,a,d), B_a(2,a,s), cause(1),
     cause(4), cause(5), preRho(1,3), preRho(4,3), preRho(5,3), cont(1,{4,5}),
      cont(4,{1,5}), cont(5,{1,4})}

    {D_a(4,a,s), B_a(2,a,d), C_a(3,a,d), A_a(1,a,s), E_a(5,a,s), cause(2),
     cause(3), preRho(2,2), preRho(3,2), cont(2,{3}), cont(3,{2})}

    {D_a(4,a,s), C_a(3,a,d), A_a(1,a,s), E_a(5,a,d), B_a(2,a,s), cause(3),
     cause(5), preRho(3,2), preRho(5,2), cont(3,{5}), cont(5,{3})}

    {D_a(4,a,s), B_a(2,a,d), C_a(3,a,s), A_a(1,a,d), E_a(5,a,s), cause(1),
     cause(2), preRho(1,2), preRho(2,2), cont(1,{2}), cont(2,{1})}
\end{verbatim} }

Cause 5, for example, appears in the first and third repairs, which are an S-repair and C-repair, resp. If we do not want to start inspecting the kinds of repairs where a cause appears, and we haven't pruned non-C-repairs,
then we may pose a
 query of the form $\nit{preRho}(5,N)?$ against this program under. This is done by inserting the query at the end of the program (in file \verb+file+), as \verb+preRho(5,N)?+.  Then in the command line, one types: \ \verb+dlv -brave file+, obtaining as expected:

{\footnotesize \begin{verbatim}
    2
    3
\end{verbatim} }

\noindent  The least of the returned values will give us the global pre-responsibility value, which can be used to compute tuple 5's (global) responsibility:  $\frac{1}{2}$. \ If we want all causes with their local pre-responsibilities, we pose instead the query: \verb+preRho(T,N)?+.

If, as in Example \ref{ex:rep1}, we impose weak constraints to obtain only C-repairs:

{\footnotesize \begin{verbatim}
    :~ A_a(T,X,d).
    :~ B_a(T,X,d).
    :~ C_a(T,X,d).
    :~ D_a(T,X,d).
    :~ E_a(T,X,d).
\end{verbatim} }

\noindent we obtain:

{\footnotesize \begin{verbatim}
    Best model: {D_a(4,a,s), B_a(2,a,d), C_a(3,a,s), A_a(1,a,d), E_a(5,a,s),
                 cause(1), cause(2), preRho(1,2), preRho(2,2), cont(1,{2}),
                 cont(2,{1})}
    Cost ([Weight:Level]): <[2:1]>

    Best model: {D_a(4,a,s), C_a(3,a,d), A_a(1,a,s), E_a(5,a,d), B_a(2,a,s),
                 cause(3), cause(5), preRho(3,2), preRho(5,2), cont(3,{5}),
                 cont(5,{3})}
    Cost ([Weight:Level]): <[2:1]>

    Best model: {D_a(4,a,s), B_a(2,a,d), C_a(3,a,d), A_a(1,a,s), E_a(5,a,s),
                 cause(2), cause(3), preRho(2,2), preRho(3,2), cont(2,{3}),
                 cont(3,{2})}
    Cost ([Weight:Level]): <[2:1]>
\end{verbatim} }

\noindent As expected for C-repairs, the local pre-responsibilities for a cause coincide.
\boxtheorem
\end{example}

\section{\ \red{An Example of Null-Based Causes with DLV}}\label{sec:B}

\begin{example} \label{ex:veryLast} \ (ex. \ref{ex:cause3} and \ref{ex:last} cont.) \ In this DLV program \verb+S+ stands for $\nit{Store}$, and \verb+R+, for $\nit{Receives}$. The tuples already contain tuple-ids. The program is similar to that in Example \ref{ex:logicP}.

{\footnotesize \begin{verbatim}
        S(1,a2).  S(2,a3).  R(3,a3,a1).  R(4,a3,a4).  R(5,a3,a5).

         S_a(T,X,t) :- S(T,X).
         S_a(T,X,t) :- S_a(T,X,u).
       R_a(T,X,Y,t) :- R(T,X,Y).
       R_a(T,X,Y,t) :- R_a(T,X,Y,u).

      S_a(T,null,u) :- S_a(T,X,t), R_a(T2,X,Y,t), X != null, not R_a(T2,null,Y,u).
    R_a(T,null,Y,u) :- R_a(T,X,Y,t), S_a(T2,X,t), X != null, not S_a(T2,null,u).

        S_a(T,X,fu) :- S_a(T,X,u), not auxS1(T,X).
         auxS1(T,X) :- S(T,X), S_a(T,null,u), X != null.

      R_a(T,X,Y,fu) :- R_a(T,X,Y,u), not auxR1(T,X,Y), not auxR2(T,X,Y).
       auxR1(T,X,Y) :- R(T,X,Y), R_a(T,null,Y,u), X != null.
       auxR2(T,X,Y) :- R(T,X,Y), R_a(T,X,null,u), Y != null.

         S_a(T,X,s) :- S_a(T,X,fu).
         S_a(T,X,s) :- S(T,X), not auxS(T).
            auxS(T) :- S_a(T,X,u).

       R_a(T,X,Y,s) :- R_a(T,X,Y,fu).
       R_a(T,X,Y,s) :- R(T,X,Y), not auxR(T).
            auxR(T) :- R_a(T,X,Y,u).
\end{verbatim} }

The query \verb+S(T1,X), R(T2,X,Y)?+  to the program under the brave semantics returns tuples of the form \ \verb+T1 X T2 Y+, showing that tuples (with tids) 2,3,4,5 are responsible for the violation of the the DC:

{\footnotesize \begin{verbatim}
    2, a3, 3, a1
    2, a3, 4, a4
    2, a3, 5, a5
\end{verbatim}}

Two stable models are returned, corresponding to two attribute-based repairs:

{\footnotesize \begin{verbatim}
{S_a(1,a2,s), S_a(2,a3,s), R_a(3,null,a1,s), R_a(5,null,a5,s), R_a(4,null,a4,s)}
{S_a(1,a2,s), R_a(3,a3,a1,s), R_a(4,a3,a4,s), R_a(5,a3,a5,s), S_a(2,null,s)}
\end{verbatim}}

\vspace{-0.5cm}\boxtheorem
\end{example}

\end{document}